\documentclass[preprint,journal]{vgtc}            

\onlineid{0}

\preprinttext{To appear in IEEE Transactions on Visualization and Computer Graphics.}

\vgtccategory{Research}

\vgtcpapertype{algorithm/technique}

\usepackage{amsfonts}

\title{Visual Analytics Using Tensor Unified Linear Comparative Analysis\vspace{-10pt}}

\author{%
  Naoki Okami, Kazuki Miyake, \authororcid{Naohisa Sakamoto}{0000-0002-9210-467X}, \authororcid{Jorji Nonaka}{0000-0001-6809-6393}, and \authororcid{Takanori Fujiwara}{0000-0002-6382-2752}
}

\authorfooter{
  \item
  	Naoki Okami, Kazuki Miyake, and Naohisa Sakamoto are with Kobe University.
  	E-mail: \{237x012x, 2165002t\}@stu.kobe-u.ac.jp, naohisa.sakamoto@people.kobe-u.ac.jp.
  \item
  	Jorji Nonaka and Kazuki Miyake (Student trainee) are with RIKEN R-CCS. 
  \item
  	Takanori Fujiwara was with Linköping University. He is now with the University of Arizona.
  	E-mail: tfujiwara@arizona.edu  	
}

\abstract{Comparing tensors and identifying their (dis)similar structures is fundamental in understanding the underlying phenomena for complex data.  Tensor decomposition methods help analysts extract tensors' essential characteristics and aid in visual analytics for tensors. In contrast to dimensionality reduction (DR) methods designed only for analyzing a matrix (i.e., second-order tensor), existing tensor decomposition methods do not support flexible comparative analysis. To address this analysis limitation, we introduce a new tensor decomposition method, named \textit{tensor unified linear comparative analysis} (TULCA), by extending its DR counterpart, ULCA, for tensor analysis. TULCA integrates discriminant analysis and contrastive learning schemes for tensor decomposition, enabling flexible comparison of tensors. We also introduce an effective method to visualize a core tensor extracted from TULCA into a set of 2D visualizations. We integrate TULCA's functionalities into a visual analytics interface to support analysts in interpreting and refining the TULCA results. We demonstrate the efficacy of TULCA and the visual analytics interface with computational evaluations and two case studies, including an analysis of log data collected from a supercomputer.
\vspace{-3pt}
}

\keywords{Tensor decomposition, tensor analysis, contrastive learning, dimensionality reduction, interpretability,  supercomputer.\vspace{-10pt}}

\teaser{
  \vspace{-3pt}
  \centering
  \includegraphics[width=\linewidth,height=0.5\linewidth]{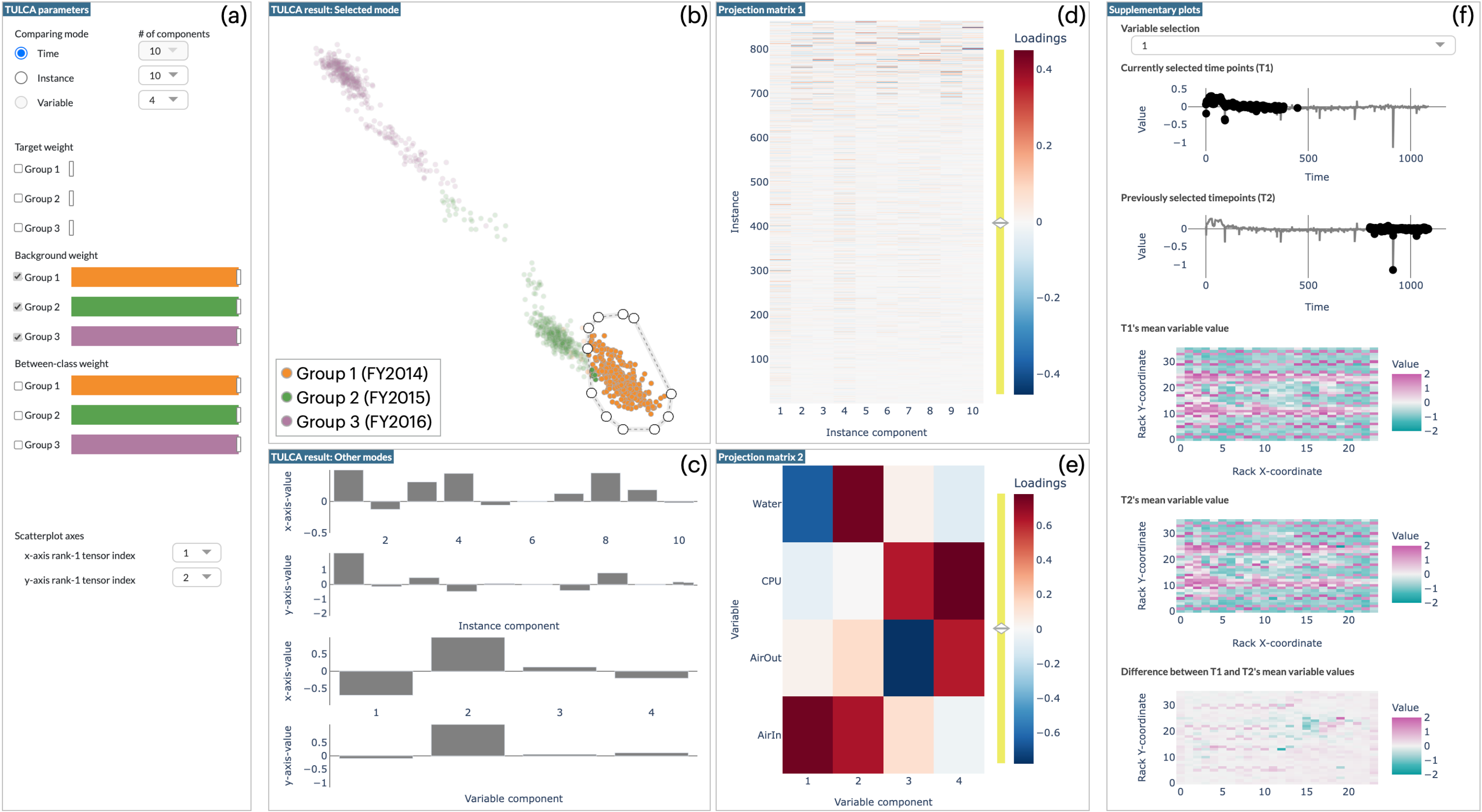}
  \caption{%
  A visual interface for analyzing supercomputer log data using \textit{tensor unified comparative analysis} (TULCA).
  The interface supports analysts in interactively (1) steering TULCA, (2) interpreting the TULCA results, and (3) examining those results in the context of the original log data. 
  To steer TULCA (a): the current parameters are set to compare timepoints and maximize their differences among Groups 1--3. 
  To interpret the TULCA results (b--e): (b) a 2D scatterplot shows how TULCA successfully separated the timepoints; 
  (c) bar charts present the instance and variable components extracted by TULCA, identifying key contributors to each scatterplot axis (e.g., the second variable component); (d, e) heatmaps illustrate how instance and variable components are derived from the original data ((d): instance, (e): variable). Dark red and blue cells, respectively, indicate high associations between the original instances/variables and their corresponding components.
  To examine the TULCA results with the original data (f), the analyst uses supplementary visualizations aligned with the selected timepoints from (b).
  \vspace{-5pt}
  }
  \label{fig:teaser}
}

\graphicspath{{figs/}{figures/}{pictures/}{images/}{./}} 

\usepackage{tabu}                      
\usepackage{booktabs}                  
\usepackage{lipsum}                    
\usepackage{mwe}                       
\usepackage{cite}

\usepackage{mathptmx}                  
\usepackage{amsmath} 
\usepackage{bm}         
\usepackage[mathscr]{euscript}   
\usepackage{amssymb}    
\usepackage{amsbsy}
\usepackage{amsfonts}

\newcommand{\tr}{\mathrm{tr}}
\DeclareMathOperator*{\argmax}{arg\,max}

\newcommand{\ModeProduct}[1]{\times_{\!#1}}

\newcommand{\Scalar}[1]{\lowercase{#1}}
\renewcommand{\Vec}[1]{\mathbf{\lowercase{#1}}}
\newcommand{\Mat}[1]{\mathbf{\uppercase{#1}}}
\newcommand{\Tensor}[1]{\pmb{\mathscr{\uppercase{#1}}}}

\newcommand{\Identity}[1]{\Mat{I}_{#1}}

\newcommand{\NClasses}{L}
\newcommand{\Class}{l}
\newcommand{\NModes}{N}
\newcommand{\Mode}{n}
\newcommand{\NTimes}{T}
\newcommand{\Time}{t}
\newcommand{\NInsts}{S}
\newcommand{\Inst}{s}
\newcommand{\NVars}{V}
\newcommand{\Var}{v}
\newcommand{\NFibers}[1]{K_{#1}}
\newcommand{\CPDecompR}{R}
\newcommand{\CPDecompIndex}{r}
\newcommand{\CPFactorMat}[1]{\Flatten{\Mat{V}}{#1}}
\newcommand{\CPFactor}[2]{\Flatten{\Vec{v}}{#1}_{#2}}

\newcommand{\Flatten}[2]{#1^{(#2)}}
\newcommand{\ProjMat}[1]{\Mat{M}_{#1}}
\newcommand{\CovWithin}[1]{\Mat{C}_{\mathrm{wi}_{#1}}}
\newcommand{\CovBetween}[1]{\Mat{C}_{\mathrm{bw}_{#1}}}
\newcommand{\CovNumerator}[1]{\Mat{C}_{\mathrm{a}_{#1}}}
\newcommand{\CovDenominator}[1]{\Mat{C}_{\mathrm{b}_{#1}}}
\newcommand{\GammaNumerator}[1]{\gamma_{\mathrm{a}_{#1}}}
\newcommand{\GammaDenominator}[1]{\gamma_{\mathrm{b}_{#1}}}

\newcommand{\TargetWeight}[1]{\Scalar{w}_{\mathrm{tg}_{#1}}}
\newcommand{\BackgroundWeight}[1]{\Scalar{w}_{\mathrm{bg}_{#1}}}
\newcommand{\BetweenWeight}[1]{\Scalar{w}_{\mathrm{bw}_{#1}}}

\newcommand{\MeanVec}[1]{\bm{\mu}_{#1}}
\newcommand{\Label}{y}

\DeclareMathAlphabet{\mathcal}{OMS}{cmsy}{m}{n}

\usepackage[font={scriptsize,sf}]{caption}
\usepackage{multicol}
\begin{document}




\firstsection{Introduction}

\maketitle

\urlstyle{rm}

\setlength{\abovedisplayskip}{2pt}
\setlength{\belowdisplayskip}{1pt}

Comparison of data groups is essential to analyze underlying factors that \textit{characterize} each group, leading to a deeper understanding of the targeted phenomena~\cite{Fujiwara:2022:ULCA,fujiwara2019supporting,abid2018exploring,hare2015using,gleicher2013explainers,yasir2015comparison,Liu:2019:TPFlow}.
For example, comparing stable and unstable operation periods of a supercomputer is necessary to identify the root cause of unstable operations.
Such comparative analysis has traditionally focused on high-dimensional data~\cite{Fujiwara:2022:ULCA}, which is often represented as a matrix.
However, comparative analysis is no longer limited to just matrices. 
Applications, such as the Internet of Things and machine learning (ML), use considerably different data structures: health monitoring devices~\cite{banos2015mhealth,wang2022iot} or supercomputer operations~\cite{ott2020oda,brandt2023oda} often produce multivariate time-series data.
Multivariate time-series data is typically modeled as a third-order tensor (or a 3D array) with time, instance, and variable axes and inevitably differs from a matrix (i.e., a second-order tensor).
The additional axis highlights the need to develop comparative analysis methods for high-order tensors.

Tensor decomposition methods~\cite{kolda2009tensor,lu2011survey,oseledets2011tensor} aid in analyzing high-order tensors.
These methods can be viewed as analogs to dimensionality reduction (DR) methods for matrices---tensor decomposition methods help analysts extract essential information from high-order tensors (e.g., dominant multivariate temporal patterns). 
The majority of tensor decomposition methods simplify high-order tensors without using group information~\cite{kolda2009tensor} (e.g., Tucker decomposition~\cite{tucker1966some}). 
More critically, they do not focus on finding patterns that are specific to group (dis)similarities.
As discussed above, identifying group-specific patterns is essential for comparative analysis, highlighting an analytical gap. 
One tensor decomposition method directly related to comparative analysis is tensor discriminant analysis (TDA)~\cite{lai2013sparse}.
However, TDA follows the discriminant analysis scheme and only uncovers differentiating factors of groups (e.g., temporal patterns that classify groups).
Only extracting differentiating factors is often insufficient when needed to satisfy diverse comparative analysis tasks.
As another essential analysis task, contrastive learning\footnote{
There is another ML scheme also called contrastive learning, which aims to learn an embedding from similar and dissimilar instance pairs to place similar instances close together in the embedding space~\cite{crl}.}~\cite{abid2018exploring,zou2013contrastive,ge2016rich} aims to extract salient factors present in one group relative to the others.
For instance, contrastive principal component analysis (cPCA) for matrices~\cite{abid2018exploring} produces an embedding space where one matrix group has a high variance but another group does not.
Finding such salient factors enriches the understanding of each group's characteristics, but tensor decomposition methods lack this aspect.
Further, as demonstrated by Fujiwara et al.~\cite{Fujiwara:2022:ULCA}, simultaneously considering these two factors (differentiating and salient factors) derives new analytical insights, such as identifying a political stance that clearly separates opinions between different political party supporters but still shows a considerable variety in each party.
Despite these research efforts, we still lack tensor decomposition methods that flexibly support various comparative analysis tasks.  

We introduce \textit{tensor unified linear comparative analysis} (\textit{TULCA}), a new tensor decomposition method that enables flexible comparison of tensor groups.
TULCA extends its DR counterpart, ULCA~\cite{Fujiwara:2022:ULCA}, for tensor analysis.
TULCA unifies two ML schemes---discriminant analysis~\cite{izenman2008modern} and contrastive learning~\cite{zou2013contrastive}---to facilitate comparisons that cannot be achieved when solely using one of the schemes.
Specifically, TULCA introduces a tensor extension of cPCA (TcPCA), integrates TcPCA with TDA, and fills the gaps among analysis targets of TcPCA and TDA.  
By filling the gaps, TULCA can not only extract factors that sufficiently differentiate groups (similar to TDA) but also simultaneously preserve a higher variance for a certain group (similar to TcPCA). 
These two abilities are critical in real-world use cases, such as analyzing supercomputer behaviors. 
Supercomputers can have stable and unstable operation periods, and operation analysts need to be able to differentiate between these two periods while still having a significant variety for the unstable periods.
For instance, the unstable period may have higher temperature of CPU cooling water in general but, at the same time, show large fluctuations in the temperature, both of which may contribute to the unstable operations.

Our core contribution is TULCA. 
Relatedly, our second contribution is a set of 2D visualizations that we design to summarize the core tensor information extracted by applying the CANDECOMP/PARAFAC (CP) decomposition~\cite{harshman1970foundations,carroll1970analysis} to the core tensor. 
These visualizations address the challenge of interpreting TULCA results.
Similar to TDA, TULCA produces a core tensor that has the same number of axes as the original tensor.
Given how the core tensor of a high-order tensor has at least three axes, it is not trivial to visualize and review the core tensor.
For example, simply visualizing all elements in an $\NModes$-order tensor requires a heatmap with $\NModes$ axes.
The CP decomposition allows us to represent the core tensor as a set of essential vectors and to produce simpler visualizations to review the core tensor.
Third, we develop an interactive visual analytics interface (\autoref{fig:teaser}) based on this set of visualizations. The interface is designed to help interpret the TULCA result and steer TULCA for analytical decision-making based on the findings.
Lastly, to demonstrate the efficacy of TULCA and its visual analytics interface, we conduct multiple computational evaluations as well as present two case studies analyzing a supercomputer log dataset and a mobile health dataset.
By analyzing supercomputer logs with a supercomputer operational staff, we identified the direct influence of the modifications to the physical cooling facility on compute rack temperatures as well as sensor failures in several racks. These findings are convincing to the end user and validate the efficacy of our approach.

We provide a demonstration video of the interface and the source code of TULCA in the supplementary material~\cite{supp}.

\vspace{-1pt}
\section{Related Work}
\vspace{-1pt}

We survey prior work about DR methods, tensor decomposition methods, and visual analytics of tensors. 

\vspace{-1pt}
\subsection{Dimensionality Reduction for Second-Order Tensors}
\label{sec:dr}

DR methods promote exploratory data analysis of high-dimensional data (modeled as a second-order tensor) and can be categorized as either linear~\cite{cunningham2015linear} or nonlinear DR~\cite{tenenbaum2000global,van2009dimensionality}. 
Linear methods, such as PCA~\cite{Jolliffe2002Principal}, produce a linear mapping from the original data to a low-dimensional representation. 
Nonlinear methods such as t-SNE~\cite{van2008visualizing} often focus on preserving local neighborhoods.
Our work is more closely related to linear DR methods, given how TULCA produces a \textit{multilinear} mapping between the original and core tensors.
Linear and multilinear mappings provide benefits in interpreting the results since the mapping explains how the results are derived from the original data.

Several linear DR methods support comparative analysis of data groups by employing either discriminant analysis or contrastive learning~\cite{izenman2008modern,zou2013contrastive}.
Discriminant analysis produces a low-dimensional representation that maximizes the separation of each group by identifying differentiating factors among groups.
Identifying such factors is accomplished by employing linear discriminant analysis (LDA)~\cite{izenman2008modern} and its variants~\cite{clemmensen2011sparse,guo2007regularized,wen2018robust}.
Contrastive learning optimizes a low-dimensional representation such that one group has a salient statistical feature relative to another group.
For example, cPCA~\cite{abid2018exploring,ge2016rich} produces a low-dimensional space where one group has a high variance but the other groups do not.
Representative linear DR methods using contrastive learning include cPCA, computationally efficient cPCA~\cite{salloum2022cpca++,golkar2023online,lu2024geometric}, and extended cPCA for different analysis targets (e.g., categorical data analysis)~\cite{fujiwara2019supporting,boileau2020exploring,fujiwara2023contrastive}.
Notably, ULCA~\cite{Fujiwara:2022:ULCA} unifies both discriminant analysis and contrastive learning schemes. 
Through this unification, ULCA not only provides the same functionalities as PCA, cPCA, and LDA but also supports flexible comparative analysis by utilizing these methods concurrently. 
However, ULCA does not directly support analyzing high-order tensors (third-order or more). 
Our work generalizes ULCA to TULCA to support comparative analysis of high-order tensors.

\subsection{Tensor Decomposition for High-Order Tensors}
\label{sec:tensor-decomposition}

Analogous to DR methods, tensor decomposition methods simplify high-order tensors and extract their essential information~\cite{pajarola2021tensor,rabanser2017introduction}.
Representative methods include Tucker decomposition~\cite{tucker1966some} and the CP decompositions~\cite{harshman1970foundations,carroll1970analysis}.
Tucker decomposition is a high-order extension of PCA and extracts a core tensor from the original tensor.
Parallel to a lower-dimensional representation of the high-dimensional data, a core tensor provides a summary of the original tensor.
The CP decomposition decomposes a tensor as the sum of a finite number of rank-one tensors.
Though Tucker and CP decompositions aid in exploratory data analysis by reducing the space to explore~\cite{Liu:2019:TPFlow}, these methods do not focus on finding patterns that are specific to group (dis)similarities. 

Several tensor decomposition methods are designed for comparative analysis.
Discriminant analysis, a type of comparative analysis, can be performed with TDA methods ~\cite{lai2013sparse,tao2007general,yan2005discriminant}, which are extensions of LDA for tensors. 
TULCA shares algorithmic similarities with TDA and we discuss the technical differences in \autoref{sec:methodology}.
However, no tensor decomposition method is designed for contrastive learning~\cite{zou2013contrastive}---another important scheme for comparative analysis.
One possible approach is to slice and convert a tensor to a single matrix and then apply DR to this matrix~\cite{Fujiwara:2021:MultiDR,kiers1991hierarchical,kiers2000towards}.
However, this approach combines multiple tensor axes into one axis (e.g., mixing temporal and variable axes).
Consequently, the resultant low-dimensional axis does not clearly distinguish different axes, and the latent structure responsible for preserving information may be heavily biased toward some influential axes and neglect others (e.g., temporal changes).
Also, research reports this approach degrades the quality of the latent structure preservation when compared to tensor decomposition methods~\cite{yang2004two,yang2005two}.
Addressing these gaps, our work introduces TULCA.
TULCA is the first tensor decomposition method that supports contrastive learning. 
Notably, TULCA can support more than contrastive learning. 
It enables analysts to flexibly switch between comparative analysis schemes (i.e., discriminant analysis or contrastive learning) or even a mixture of both schemes. 

\subsection{Interactive Visual Analysis of Tensors}

Tensors examined in visual analytics are multivariate time-series data or multivariate spatiotemporal data~\cite{andrienko2003exploratory,andrienko2013visual}.
Such data is often modeled as a third-order tensor of timepoints, instances/locations, and variables.
They may be even modeled as a higher-order tensor when analysts separate the time axis to consider periodic temporal behaviors (e.g., hours of the day, days of the week).
Visual analytics of high-order tensors have two notable challenges. 
First, analysts need to reviewing data patterns related to multiple axes (e.g., spatiotemporal patterns).
Second, high-order tensors encompass a large number of elements (e.g., 1000 timepoints, 100 instances, and 10 variables produce 1,000,000 elements).
To identify interesting patterns from such complex, large data, visual analytics research enhances tensor decomposition methods.

For instance, Voila~\cite{cao2018voila} introduces an extension of Tucker decomposition that assists analysts detecting visual anomaly from streaming multivariate spatiotemporal data.
TPFlow~\cite{Liu:2019:TPFlow} enhances the CP decomposition by partitioning a tensor to sub-tensors based on their similarities and then applying tensor decomposition to each sub-tensor such that decomposed results better capture the data variations.
Another approach is to apply multiple DR methods to separately summarize each axis of the tensors~\cite{Fujiwara:2021:MultiDR,fujita2022visual}.
For example, MulTiDR~\cite{Fujiwara:2021:MultiDR} first uses PCA to remodel a third-order tensor as a second-order tensor and then uses UMAP to overview the second-order tensor with a 2D scatterplot. 
Our work is more closely related to Viola and TPFlow, both of which use tensor decomposition for visual analytics~\cite{cao2018voila,Liu:2019:TPFlow}. 
However, while Viola and TPFlow produce a general summary of a tensor, TULCA is designed to comparatively analyze tensor groups, such as identifying differentiating and salient factors of each group.

\begin{table}[t]
    \renewcommand{\arraystretch}{0.75}
    \setlength{\tabcolsep}{3pt}
    \centering
    \caption{Summary of notation.\vspace{-2pt}}
    \label{tab:notations}
    \scriptsize
    \begin{tabular}{rl}
        \toprule
            $\Tensor{X}$, $\Tensor{Z}$ & Original and core tensors \\
            $\NModes$, $\NClasses$, $\NInsts$, $\NVars$, $\NTimes$ & \#s of modes, classes, instances, variables, timepoints\\
            $\Mode$, $\Class$, $\Inst$, $\Var$, $\Time$ & Indices for modes, classes, instances, variables, timepoints\\
            $\smash{\NFibers{\Mode}}$ & Length of mode-$\Mode$\\
            $\Flatten{\Mat{X}}{\Mode}$ & Mode-$\Mode$ matricized tensor $\Tensor{X}$\\
            $\smash{\ProjMat{\Mode}}$ & Projection matrix for mode-$\Mode$\\
            $\smash{\ModeProduct{\Mode}}$ & $\Mode$-mode product operator\\
            $\Flatten{\CovWithin{}}{\Mode}, \Flatten{\CovBetween{}}{\Mode}$ & Within-, between-class covariance matrices for mode-$\Mode$\\
            $\smash{\TargetWeight{}}$, $\smash{\BackgroundWeight{}}$, $\smash{\BetweenWeight{}}$ & Weights for target, background, between-class covariances\\
            $\CPFactorMat{\Mode}$ & Factor matrix of mode-$\Mode$ generated by the CP decomposition\\
        \bottomrule
    \end{tabular}
\end{table}

\vspace{-1pt}
\section{Methodology}
\label{sec:methodology}

We introduce TULCA, a tensor decomposition method that can flexibly support two comparative analysis schemes---discriminant analysis and contrastive learning---or even a mixture of both schemes.
We first provide the necessary background of high-order tensors and TDA.
TULCA embraces the analytical functionality of TDA, and resultingly TULCA also shares similarities with TDA's optimization problem. 
Thus, understanding the fundamentals of TDA can help better understand TULCA's design.
Lastly, we discuss how we design TULCA. 

\vspace{-1pt}
\subsection{Background}

We describe the basics of high-order tensors and the notation used in this work. 
We then provide a brief introduction of TDA.

\vspace{-1pt}
\subsubsection{High-Order Tensors, Notation, and Matricization}
\label{sec:matricization}

\textbf{High-order tensors.} An $\smash{\NModes}$-order tensor can be considered as data stored in an $\smash{\NModes}$-dimensional array.
In this work, we denote a tensor to be high-order when $\smash{\NModes \geq 3}$.
This definition distinguishes a high-order tensor from a vector (i.e., $\smash{\NModes {=} 1}$) and a matrix (i.e., $\smash{\NModes {=} 2}$). 
Each axis of a tensor is called \textit{mode}.
For example, multivariate time-series data can be modeled as a third-order tensor with time, instance, and variable modes.
Let $\smash{\NFibers{\Mode}}$ denote the length of mode-$\smash{\Mode}$.
Then, an $\smash{\NModes}$-order tensor can be represented as \scalebox{0.95}{$\smash{\Tensor{X} \in \mathbb{R}^{\NFibers{1} \times \NFibers{2} \times \cdots \times \NFibers{\NModes}}}$}.
In this section, we use a third-order tensor for graphical explanations of tensor operations but these operations can be applied to any $\smash{\NModes}$-order tensor.

\textbf{Notation.} We follow the notation conventions~\cite{kolda2009tensor}.
Scalars are denoted by italic letters (e.g., $\smash{\Scalar{x}}$, $\smash{\NModes}$).
Vectors, matrices, and tensors are, respectively, denoted with boldface lowercase (e.g., $\smash{\Vec{x}}$), boldface uppercase (e.g., $\smash{\Mat{x}}$), and boldface Euler script (e.g., $\smash{\Tensor{x}}$).
Indices use lowercase letters and range from 1 to their uppercase counterparts.
For example, we represent time, instance, variable indices, respectively, with $\smash{\Time = 1, \cdots, \NTimes}$; $\smash{\Inst = 1, \ldots, \NInsts}$; and $\smash{\Var = 1, \ldots, \NVars}$.
An index for modes is $\smash{\Mode = 1, \cdots, \NModes}$.
\autoref{tab:notations} summarizes the notation.

\begin{figure}[tb]
    \centering
    \includegraphics[width=0.95\linewidth]{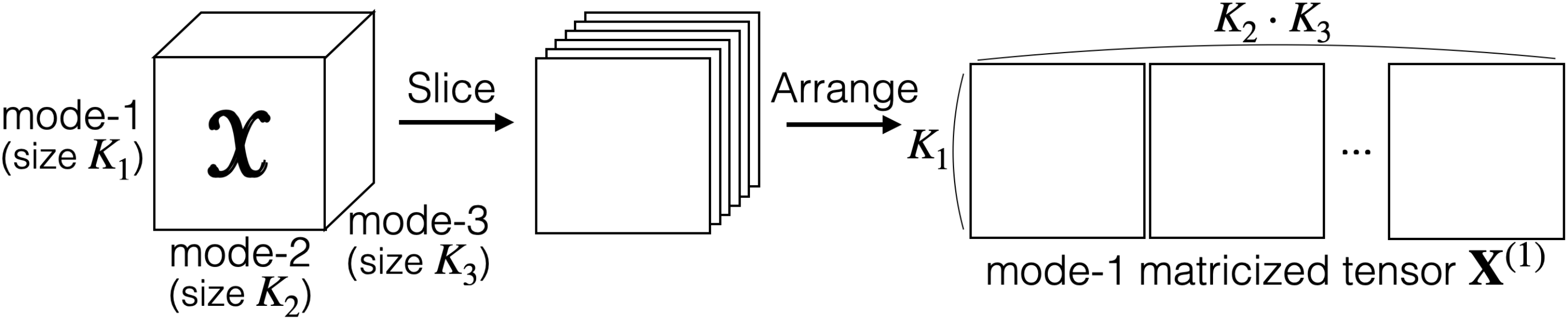}
    \caption{Mode-$1$ matricization of a third-order tensor.}
    \label{fig:matricization}
\end{figure}

\textbf{Matricization of tensors.} 
We frequently apply matricization~\cite{kolda2009tensor} (also known as unfolding and flattening) to tensors.
Matricization slices an $\smash{\NModes}$-order tensor along one mode and reorders the elements in each slice to produce a matrix, as shown in \autoref{fig:matricization}. 
For example, matricization can rearrange a $\smash{\NTimes \times \NInsts \times \NVars}$ tensor as a matrix with the size of $\smash{\NTimes \times \NInsts \NVars}$ (sliced along the time mode), $\smash{\NInsts \times \NTimes \NVars}$ (sliced along the instance mode), or $\smash{\NVars \times \NTimes \NInsts}$ (sliced along the variable mode).
We denote the mode-$\smash{\Mode}$ matricization result of $\Tensor{X}$ as $\smash{\Flatten{\Mat{X}}{\Mode}}$ where $\smash{\Flatten{\Mat{X}}{\Mode} \in \mathbb{R}^{\NFibers{\Mode} \times \prod_{m \neq n}\NFibers{m}}}$. 
We refer readers to Kolda and Bader's work~\cite{kolda2009tensor} for a more detailed overview.

\begin{figure}[t]
    \centering
    \includegraphics[width=0.95\linewidth]{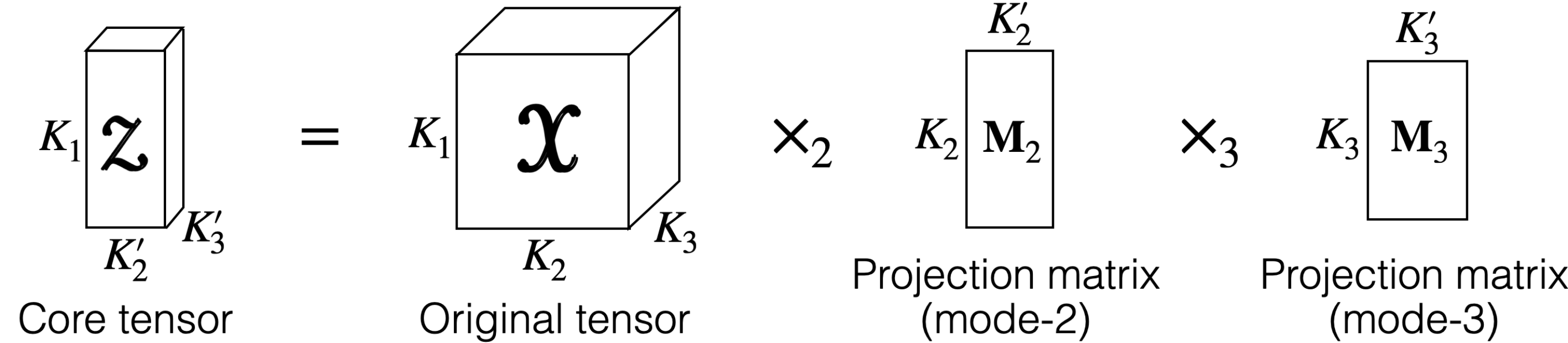}
    \caption{Core tensor extraction from a third-order tensor.}
    \label{fig:mode-product}
\end{figure}

\subsubsection{Tensor Discriminant Analysis (TDA)}
\label{sec:TDA}

TDA generalizes LDA~\cite{izenman2008modern} for high-order tensors. 
We first explain LDA and discuss how it is extended for TDA.

\textbf{Linear discriminant analysis (LDA).}
LDA~\cite{izenman2008modern} embeds a matrix, $\smash{\Mat{X} \in \mathbb{R}^{\NInsts \times \NVars}}$, into a low-dimensional representation, $\smash{\Mat{Z} \in \mathbb{R}^{\NInsts \times \NVars'} (\NVars' \leq \NVars)}$, to maximize separation among groups.
To do so, LDA finds a projection matrix, $\smash{\ProjMat{} \in \mathbb{R}^{\NVars \times \NVars'}}\!\!$, such that the representation, $\smash{\Mat{Z} {=} \Mat{X} \ProjMat{}}$, maximizes the distance between each group's centroid (\textit{between-class} variance) while minimizing variance within each group (\textit{within-class} variance).
More precisely, LDA performs the following optimization:
\begin{equation}
    \ProjMat{} = 
   \argmax_{\ProjMat{}^\top \ProjMat{} = \Identity{\NVars'}} ~ \frac{\tr(\ProjMat{}^\top \CovBetween{} \ProjMat{})}{\tr(\ProjMat{}^\top \CovWithin{} \ProjMat{})}
   \label{eq:lda}
\end{equation}
where $\smash{\Identity{\NVars'}}$ is the identify matrix of size $\smash{\NVars'}$. $\smash{\CovWithin{}}$ and $\smash{\CovBetween{}}$ are the within-class and between-class covariance matrices:
\begin{equation}
   \CovWithin{} = \frac{1}{\NInsts} \sum_{\Inst=1}^{\NInsts} (\Vec{x}_\Inst - \MeanVec{\Label_\Inst})(\Vec{x}_\Inst - \MeanVec{\Label_\Inst})^\top
   \label{eq:cov-within-lda}
   \vspace{-3pt}
\end{equation}
\begin{equation}
   \CovBetween{} = \frac{1}{\NInsts} \sum_{\Inst=1}^{\NInsts} (\MeanVec{\Label_\Inst} - \MeanVec{})(\MeanVec{\Label_\Inst} - \MeanVec{})^\top
   \vspace{-7pt}
\end{equation}
where $\smash{\smash{\Vec{x}_\Inst \in \mathbb{R}^\NVars}}$ is the $\smash{\Inst}$-th instance, $\smash{\MeanVec{} \in \mathbb{R}^\NVars}$ is the column means of $\Mat{X}$, and $\smash{\MeanVec{\Label_\Inst} \in \mathbb{R}^\NVars}$ is the column means of group instances that $\smash{\Vec{x}_\Inst}$ belongs to.
Here, $\smash{\Label_\Inst \in \{1, \ldots, \NClasses\}}$ ($\NClasses$: the number of classes or groups) is the $\Inst$-th element of $\Vec{y}$---a vector of group label information for each instance.

\textbf{Extension of LDA to TDA.}
We now handle an $\smash{\NModes}$-order tensor, $\smash{\Tensor{X} \in \mathbb{R}^{\NFibers{1} \times \NFibers{2} \times \cdots \times \NFibers{\NModes}}}$ ($\smash{\NFibers{\Mode}}$ is the length of mode-$\smash{\Mode}$).
Like LDA, TDA~\cite{lai2013sparse} aims to maximize the separation among groups in a resultant core tensor, $\smash{\Tensor{Z} \in \mathbb{R}^{\NFibers{1} \times \NFibers{2}' \times \cdots \times \NFibers{\NModes}'}}$ ($\smash{\NFibers{n}' \leq  \NFibers{n}}$ for $\smash{n \geq 2}$).
We call each dimension in each compressed mode \textit{component} (i.e., in $\smash{\Tensor{Z}}$, mode-$\smash{\Mode}$ has $\smash{\NFibers{n}'}$ components).  
For simplicity, we make two assumptions. 
First, TDA reduces the dimensionality of each mode except for the first mode (mode-$1$). 
Second, groups are defined along the first mode.
These assumptions are consistent with LDA, where LDA reduces the dimensionality of the variable mode (mode-$2$) while referring to groups of the instance mode (mode-$1$). 
TDA extracts the core tensor $\smash{\Tensor{Z}}$ in two steps. 
It first finds a projection matrix for each mode, $\smash{\ProjMat{\Mode}}$ ($\smash{n \in \{2, 3, \cdots, \NModes\}}$). 
Then, TDA performs the following matrix operation (illustrated in \autoref{fig:mode-product}):
\begin{equation}
    \Tensor{Z} = \Tensor{X} \ModeProduct{2} \ProjMat{2} \ModeProduct{3} \ProjMat{3} \cdots \ModeProduct{\NModes} \ProjMat{\NModes}
    \vspace{2pt}
    \label{eq:to-core-tensor}
\end{equation}
where $\ModeProduct{\Mode}$ computes the $\Mode$-mode product~\cite{kolda2009tensor} of a tensor and a matrix.
The $\Mode$-mode product multiplies a tensor with a matrix in mode-$\Mode$ (i.e., \scalebox{0.93}{$\smash{\Tensor{Z} = \Tensor{X} \ModeProduct{n} \ProjMat{n} \Leftrightarrow \Flatten{\Mat{Z}}{n} = \Flatten{\Mat{X}}{n} \ProjMat{n}}$}).

TDA generates $\ProjMat{\Mode}$, a projection matrix for mode-$\Mode$, to maximize the separation of mode-$1$ groups by solving the optimization below:
\begin{equation}
    \ProjMat{\Mode} = \argmax_{\ProjMat{\Mode}^\top \ProjMat{\Mode} = \Identity{\NFibers{\Mode}'}} ~ \frac{\tr \big( \ProjMat{\Mode}^\top \Flatten{\CovBetween{}}{\Mode} \ProjMat{\Mode} \big)}{\tr \big( \ProjMat{\Mode}^\top \Flatten{\CovWithin{}}{\Mode} \ProjMat{\Mode} \big)}
    \label{eq:tda}
\end{equation}
where $\smash{\Flatten{\CovWithin{}}{\Mode}}$ and \scalebox{0.9}{$\smash{\Flatten{\CovBetween{}}{\Mode}}$} are mode-$\Mode$ within-class and between-class covariance matrices, respectively.
TDA defines these matrices as:
\begin{equation}
    \Flatten{\CovWithin{}}{\Mode} = \frac{1}{\NFibers{1}} \sum_{i=1}^{\NFibers{1}} \big(\Flatten{\Mat{X}}{\Mode}_i - \Flatten{\MeanVec{{\Label_i}}}{\Mode} \big) \big( \Flatten{\Mat{X}}{\Mode}_i - \Flatten{\MeanVec{{\Label_i}}}{\Mode} \big)^\top
    \label{eq:cov-within-nmode}
    \vspace{-1pt}
\end{equation}
\begin{equation}
    \Flatten{\CovBetween{}}{\Mode} = \frac{1}{\NFibers{1}} \sum_{i=1}^{\NFibers{1}} \big( \Flatten{\MeanVec{{\Label_i}}}{\Mode} - \Flatten{\MeanVec{}}{\Mode} \big) \big( \Flatten{\MeanVec{{\Label_i}}}{\Mode} - \Flatten{\MeanVec{}}{\Mode} \big)^\top
    \label{eq:cov-between-nmode}
    \vspace{-3pt}
\end{equation}
where $\smash{\Flatten{\Mat{X}}{\Mode}_i}$ is the mode-$\Mode$ matricized tensor of the $i$-th slice of $\smash{\Tensor{X}}$ along mode-$1$. 
$\smash{\Flatten{\MeanVec{}}{\Mode}}$ is the column means of $\smash{\Flatten{\Mat{X}}{\Mode}}$.
$\smash{\Flatten{\MeanVec{{\Label_i}}}{\Mode}}$ is the column means of a group that $i$-th slice belongs to.
Here, $\smash{\Label_i \in \{1, \ldots, \NClasses\}}$ ($\NClasses$: the number of classes or groups) informs the $i$-th slice's belonging group.

\vspace{-1pt}
\subsection{Tensor Unified Linear Comparative Analysis (TULCA)}
\label{sec:TULCA}
\vspace{-1pt}

Inspired by how TDA is extended from LDA, we design TULCA to generalize ULCA~\cite{Fujiwara:2022:ULCA} for high-order tensors. 

\vspace{-2pt}
\subsubsection{Unified Linear Comparative Analysis (ULCA)}
\label{sec:ulca}
\vspace{-2pt}

ULCA~\cite{Fujiwara:2022:ULCA} supports both discriminant analysis and contrastive learning as shown in its optimization, extracting a projection matrix, $\ProjMat{}$:
\begin{equation}
    \ProjMat{} = \argmax_{\ProjMat{}^\top \ProjMat{} = \Identity{\NVars'}} \frac{ \tr ( \ProjMat{}^\top \CovNumerator{} \ProjMat{} )}
    {\tr ( \ProjMat{}^\top \CovDenominator{} \ProjMat{} ) }
    \label{eq:ulca}
    \vspace{-3pt}
\end{equation}
\begin{equation}
    \CovNumerator{} = \sum_{\Class=1}^{\NClasses} \TargetWeight{\Class} \CovWithin{\Class} + \sum_{\Class=1}^{\NClasses} \BetweenWeight{\Class} \CovBetween{\Class} + \GammaNumerator{} \Identity{\NVars}
    \vspace{-7pt}
\end{equation}
\begin{equation}
    \CovDenominator{} = \sum_{\Class=1}^{\NClasses} \BackgroundWeight{\Class} \CovWithin{\Class} +  \GammaDenominator{} \Identity{\NVars}
    \vspace{-5pt}
\end{equation}
where $\smash{\TargetWeight{\Class}}$, $\smash{\BackgroundWeight{\Class}}$, $\smash{\BetweenWeight{\Class}}$ are target, background, between-class weight parameters for group $\Class$; $\smash{\GammaNumerator{}}\!$ and $\smash{\GammaDenominator{}}\!$ are regularization parameters; $\smash{\CovWithin{\Class}}\!\!$ and $\smash{\CovBetween{\Class}}\!\!$ are within-class and between-class covariance matrices of instances belonging to group $\Class$.
More precisely, let $\smash{\delta_{\Label_s}^{\scriptscriptstyle \Class}}$ be a function judging whether instance $s$ belongs to group $\Class$ (i.e., $\smash{\delta_{\Label_\Inst}^{\scriptscriptstyle \Class} = 1}$ when $\smash{\Label_\Inst = \Class}$, otherwise $\smash{\delta_{\Label_\Inst}^{\scriptscriptstyle \Class} = 0}$). 
Then,  
\scalebox{0.9}{$
\CovWithin{\Class} = \sum_{\Inst} \delta_{\Label_\Inst}^{\scriptscriptstyle \Class} (\Vec{x}_\Inst - \MeanVec{\Label_\Inst})(\Vec{x}_\Inst - \MeanVec{\Label_\Inst})^\top / \sum_{\Inst} \delta_{\Label_\Inst}^{\scriptscriptstyle \Class}
$}
and 
\scalebox{0.9}{$
\CovBetween{\Class} = 
\sum_{\Inst} \delta_{\Label_\Inst}^{\scriptscriptstyle \Class} (\MeanVec{\Label_\Inst} - \MeanVec{})(\MeanVec{\Label_\Inst} - \MeanVec{})^\top / \sum_{\Inst} \delta_{\Label_\Inst}^{\scriptscriptstyle \Class}
$}.
The regularization parameters of ULCA uses $ \GammaNumerator{} = 0$ and $ \GammaDenominator{} = 0$ by default. 
However, $\GammaNumerator{} = 1$ automatically when $\sum_{\Class} \TargetWeight{\Class} \CovWithin{\Class} + \sum_{\Class} \BetweenWeight{\Class} \CovBetween{\Class} = 0$, and $\GammaDenominator{} = 1$ when $\sum_{\Class} \BackgroundWeight{\Class} \CovWithin{\Class} {=} 0$.
This setting is to handle either case where (1) all $\smash{\TargetWeight{\Class} = 0}$ and $\smash{\BackgroundWeight{\Class} = 0}$ for all groups or (2) all $\smash{\BackgroundWeight{\Class} = 0}$ for all groups. 
Furthermore, as in regularized LDA~\cite{guo2007regularized}, $\smash{\GammaNumerator{}}$ and $\smash{\GammaDenominator{}}$ can be manually set to add regularization terms to handle cases where $\smash{\NInsts \ll \NVars}$ (i.e., compared to the data size, limited instances are available).
Refer to Fujiwara et al.'s work~\cite{Fujiwara:2022:ULCA} for more details of $\smash{\GammaNumerator{}}$ and $\smash{\GammaDenominator{}}$.

Weight parameters, $\smash{\TargetWeight{\Class}}$, $\smash{\BetweenWeight{\Class}}$, and $\smash{\BackgroundWeight{\Class}}$ ($\smash{0 \leq \TargetWeight{\Class}, \BetweenWeight{\Class}, \BackgroundWeight{\Class} \leq 1}$) control how strongly ULCA separates each group (i.e., discriminant analysis) and preserves or eliminates each group's variance (i.e., contrastive learning).
ULCA allows analysts to flexibly conduct comparative analysis by adjusting these parameters.
For example, when $\smash{\TargetWeight{\Class} {=} 0}$, $\smash{\BetweenWeight{\Class} {=} 1}$, $\smash{\BackgroundWeight{\Class} {=} 1}$ for all groups, ULCA maximizes the between-class variances while minimizing the within-class variances. 
This configuration result is equivalent to LDA.
On the other hand, when $\smash{\BetweenWeight{\Class} {=} 0}$ for all groups, ULCA is equivalent to generalized cPCA~\cite{Fujiwara:2022:ULCA} and focuses on performing contrastive learning.
For example, if $\smash{\BetweenWeight{\Class} {=} 0}$ for all groups, $\smash{\Mat{\TargetWeight{}} {=} (1, 0, \ldots, 0)}$ and  $\smash{\Mat{\BackgroundWeight{}} {=} (0, 1, \ldots, 1)}$,
ULCA maximizes the preservation of group $1$'s variance while minimizing the other groups' variances.  
This configuration corresponds to cPCA by identifying salient factors in group $1$ relative to others.

\vspace{-2pt}
\subsubsection{Extension from ULCA to TULCA}
\label{sec:ulca-to-tulca}
\vspace{-2pt}

Our goal with TULCA is to find each mode's projection matrix, $\ProjMat{\Mode}$ ($\Mode \in \{2, 3, \cdots, \NModes\}$), that supports both discriminant and contrastive learning schemes for a high-order tensor. 
To extract a core tensor from a high-order tensor, TULCA finds $\ProjMat{\Mode}$ by solving the following:
\begin{equation}
    \ProjMat{\Mode} = \argmax_{\ProjMat{\Mode}^\top \ProjMat{\Mode} = \Identity{\NFibers{\Mode}'}} \frac{ \tr ( \ProjMat{\Mode}^\top \Flatten{\CovNumerator{}}{\Mode} \ProjMat{\Mode} )}
    {\tr ( \ProjMat{\Mode}^\top \Flatten{\CovDenominator{}}{\Mode} \ProjMat{\Mode} ) }
    \label{eq:tulca}
    \vspace{-2pt}
\end{equation}
\begin{equation}
    \Flatten{\CovNumerator{}}{\Mode} = \sum_{\Class=1}^{\NClasses} \Flatten{\TargetWeight{\Class}}{\Mode} \Flatten{\CovWithin{\Class}}{\Mode} + \sum_{\Class=1}^{\NClasses} \Flatten{\BetweenWeight{\Class}}{\Mode} \Flatten{\CovBetween{\Class}}{\Mode} + \Flatten{\GammaNumerator{}}{\Mode} \Identity{\NFibers{\Mode}}
    \label{eq:tulca-cov-numerator}
    \vspace{-6pt}
\end{equation}
\begin{equation}
    \Flatten{\CovDenominator{}}{\Mode} = \sum_{\Class=1}^{\NClasses} \Flatten{\BackgroundWeight{\Class}}{\Mode} \Flatten{\CovWithin{\Class}}{\Mode}  + \Flatten{\GammaDenominator{}}{\Mode} \Identity{\NFibers{\Mode}}
    \label{eq:tulca-cov-denominator}
    \vspace{-4pt}
\end{equation}
\scalebox{0.9}{$\Flatten{\CovWithin{\Class}}{\Mode}$}, \scalebox{0.9}{$\smash{\Flatten{\CovBetween{\Class}}{\Mode}}$}, \scalebox{0.9}{$\smash{\Flatten{\TargetWeight{\Class}}{\Mode}}$}, \scalebox{0.9}{$\smash{\Flatten{\BackgroundWeight{\Class}}{\Mode}}$}, \scalebox{0.9}{$\smash{\Flatten{\BetweenWeight{\Class}}{\Mode}}$}, \scalebox{0.9}{$\smash{\Flatten{\GammaNumerator{}}{\Mode}}$}, \scalebox{0.9}{$\smash{\Flatten{\GammaDenominator{}}{\Mode}}$} have the same role as their counterparts in ULCA (cf. \autoref{sec:ulca}).
The superscript $(\Mode)$ indicates usage for mode-$\Mode$ as in TDA (cf. \autoref{eq:tda}, \autoref{eq:cov-within-nmode}, \autoref{eq:cov-between-nmode}).
The weight parameters, $\smash{\TargetWeight{\Class}}$, $\smash{\BackgroundWeight{\Class}}$, $\smash{\BetweenWeight{\Class}}$, are the main parameters analysts should adjust based on their analysis needs.
The regularization parameters are set automatically as in ULCA unless analysts manually add strong regularization to deal with cases where $\smash{\NFibers{1}}$ is smaller relative to the data size.
After obtaining a set of projection matrices, TULCA performs the same operation as \autoref{eq:to-core-tensor} to produce a core tensor, $\smash{\Tensor{Z}}$.

By adjusting the weight parameters, analysts can control how strongly TULCA separates groups and/or preserves/eliminates information related to each group's variance.
However, unlike ULCA, adjustable weight parameters can be numerous.
An $\NModes$-order tensor with $\NClasses$ classes have $3 \NModes\NClasses$ parameters (e.g., 90 when $\NModes{=}3$ and $\NClasses{=}10$).
To reduce this parameter setting complexity, by default, TULCA uses the same weight values for each mode, i.e., \scalebox{0.9}{$\smash{\TargetWeight{\Class} = \Flatten{\TargetWeight{\Class}}{2} = \cdots = \Flatten{\TargetWeight{\Class}}{\NModes}}$}; \scalebox{0.9}{$\BackgroundWeight{\Class} = \Flatten{\BackgroundWeight{\Class}}{2} = \cdots = \Flatten{\BackgroundWeight{\Class}}{\NModes}$}; \scalebox{0.9}{$\BetweenWeight{\Class}= \Flatten{\BetweenWeight{\Class}}{2} = \cdots = \Flatten{\BetweenWeight{\Class}}{\NModes}$}.

Aiming to support high-order tensors, TULCA can be equivalent to TDA or a high-order tensor version of cPCA when using the same weight parameters as ULCA (cf. \autoref{sec:ulca}).
For example, when $\smash{\Vec{\TargetWeight{}} = (1, 0, \cdots, 0)}$, $\smash{\Vec{\BackgroundWeight{}} = (0, 1, \cdots, 1)}$,  $\smash{\Vec{\BetweenWeight{}} = (0, 0, \cdots, 0)}$, TULCA maximally preserves the variance of group $1$ while minimizing the variances of the other groups (i.e., contrasting group 1's variance with the different groups' variances).
This contrastive learning for a high-order tensor is a novel functionality introduced by TULCA, which we coin as a tensor extension of cPCA (or TcPCA).
Note that, due to the ratio optimization form (\autoref{eq:tulca}), when $\smash{\Flatten{\GammaNumerator{}}{\Mode}}$ and $\smash{\Flatten{\GammaDenominator{}}{\Mode}}$ are set to $0$, multiple sets of weight parameter values can yield the same optimization result  (e.g., $\smash{\Vec{\TargetWeight{}} = (1, \cdots, 1)}$, $\smash{\Vec{\BackgroundWeight{}} = (0, \cdots, 0)}$,  $\smash{\Vec{\BetweenWeight{}} = (1, \cdots, 1)}$ vs. $\smash{\Vec{\TargetWeight{}} = (0.1, \cdots, 0.1)}$, $\smash{\Vec{\BackgroundWeight{}} = (0, \cdots, 0)}$,  $\smash{\Vec{\BetweenWeight{}} = (0.1, \cdots, 0.1)}$).

\subsubsection{Computational Optimization of TULCA}
\label{sec:computational-optimization}
\textbf{Optimization problem solver.} To solve \autoref{eq:tulca}, we utilize ULCA's optimization methods~\cite{Fujiwara:2022:ULCA}.
They designed two different methods: (1) iterative eigenvalue decomposition and (2) manifold optimization.
While TULCA supports both methods, we apply the iterative eigenvalue decomposition due to its computational efficiency. 
For the detailed differences of the two optimizations, refer to Fujiwara et al.'s work~\cite{Fujiwara:2022:ULCA}.

\textbf{Time complexity analysis.} 
TULCA has two major calculations: the covariance matrix calculation for \scalebox{0.9}{$\smash{\Flatten{\CovWithin{\Class}}{\Mode}}$} and iterative eigenvalue decomposition to solve \autoref{eq:tulca}.
The time complexity to compute \scalebox{0.9}{$\smash{\Flatten{\CovWithin{\Class}}{\Mode}}$} for all groups is $\smash{\mathcal{O}(\sum_\Mode \NFibers{\Mode} \prod_m \NFibers{m})}$.
The iterative eigenvalue decomposition has $\smash{\mathcal{O}(\sum_\Mode \NFibers{\Mode}^3)}$.
These time complexities imply that when the number of modes, $\NModes$, is large, the covariance matrix calculation is likely to be a bottleneck (as such a tensor often satisfies \scalebox{0.9}{$\smash{\prod_m \NFibers{m} > \NFibers{\Mode}^2}$} for any mode-$\Mode$).

\textbf{Efficient optimization for parameter updates.}
We want to efficiently update the TULCA results when analysts interactively adjust the weight parameters.
Based on the time complexity, the covariance matrix calculation is the performance bottleneck.
However, as seen in \autoref{eq:tulca-cov-numerator} and \autoref{eq:tulca-cov-denominator}, the covariance matrices (\scalebox{0.9}{$\smash{\Flatten{\CovWithin{\Class}}{\Mode}}$}, \scalebox{0.9}{$\smash{\Flatten{\CovBetween{\Class}}{\Mode}}$}) do not change even if when weight parameters are updated.
Therefore, for the interactive use, TULCA pre-computes \scalebox{0.9}{$\smash{{\Flatten{\CovWithin{\Class}}{\Mode}}}$} and \scalebox{0.9}{$\smash{\Flatten{\CovBetween{\Class}}{\Mode}}$} from a given tensor.
When the weight parameters are adjusted, TULCA re-computes \scalebox{0.9}{$\smash{\Flatten{\CovNumerator{}}{\Mode}}$} and \scalebox{0.9}{$\smash{\Flatten{\CovDenominator{}}{\Mode}}$} from the pre-computed covariance matrices and then performs the iterative eigenvalue decomposition to solve \autoref{eq:tulca}.
Thus, updating the weight parameter only involves $\smash{\mathcal{O}(\sum_\Mode \NFibers{\Mode}^3)}$ computation.

\textbf{Implementation.}
We implemented TULCA with Python 3.
We used NumPy/SciPy~\cite{virtanen2020scipy} and TensorLy~\cite{JMLR:v20:18-277} for matrix and tensor calculations.
NumPy/SciPy is also used to perform the iterative eigenvalue decomposition.
We used  Pymanopt~\cite{townsend2016pymanopt} for the manifold optimization.

\section{Visual Analytics Workflow Using TULCA}

This section describes a human-in-the-loop visual analytics workflow using TULCA (\autoref{fig:workflow}) and a visual analytics interface that depicts essential information for analysis and refinement of the TULCA results.

\subsection{Essential Information for Analysis of TULCA Results}
\label{sec:essential-info}

\begin{figure}[t]
    \centering
    \includegraphics[width=0.9\linewidth]{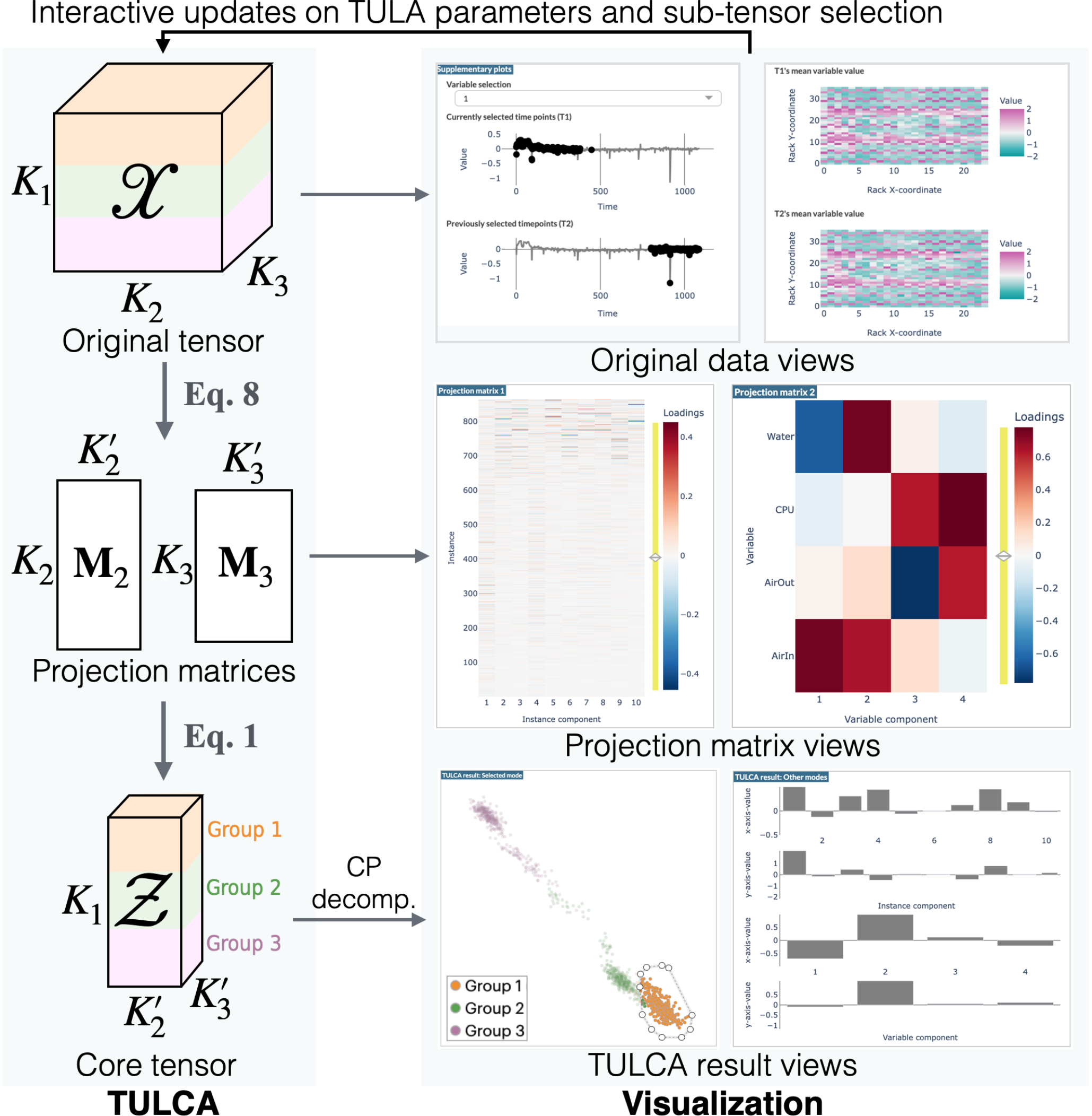}
    \caption{A visual analytics workflow using TULCA. 
    Based on \textit{user-set parameters}, TULCA generates \textit{projection matrices} and a \textit{core tensor} from the \textit{original tensor}.
    Our interface visualizes the corresponding information.
    With the interface, analysts can interactively update TULCA and the visualizations by adjusting TULCA parameters (\autoref{fig:teaser}-a), sub-tensor selection (\autoref{fig:teaser}-b), and filtering of matrix values (\autoref{fig:teaser}-d).}
    \label{fig:workflow}
\end{figure}

As discussed in \autoref{sec:ulca-to-tulca} and shown in \autoref{fig:workflow}, TULCA extracts a core tensor, $\smash{\Tensor{Z}}$, from a tensor, $\smash{\Tensor{X}}$, by applying projection matrices, $\smash{\ProjMat{\Mode}}$ ($\smash{\Mode \in \{2, \cdots, \NModes \}}$), that reflect the parameters, $\smash{\TargetWeight{\Class}}$, $\smash{\BackgroundWeight{\Class}}$, $\smash{\BetweenWeight{\Class}}$ ($\smash{\Class \in \{1, \cdots, \NClasses \}}$).
All of these are essential to analyze TULCA results.

First, \textbf{(1) the user-set parameters} reflect the underlying notion of what an analyst wants to find (e.g., differentiating factors of two groups). These parameters should be compared with the TULCA results to judge whether TULCA successfully identified the targeted factors.
Second, \textbf{(2) the core tensor} is the main output of TULCA, which contains patterns TULCA extracted (e.g., subclusters in a group). 
Third, \textbf{(3) the projection matrices} inform the numerical mapping between the original and core tensors.
Similar to a projection matrix in linear DR (e.g., PC loadings in PCA), this mapping is critical to interpreting the core tensor (e.g., variable 1 highly associates with the core tensor). 
Lastly, \textbf{(4) the original tensor} is important to contextualize the TULCA results with the information of the original data. 

We design visualizations corresponding to these four elements and integrate them into a visual analytics interface (\autoref{sec:system}).
However, visualizing the core tensor is more challenging than visualizing the other elements given how the core tensor represents $\NModes$ modes.
We describe our design process of visualizing a core tensor in \autoref{sec:core-tensor-vis}. 

\begin{figure}[t]
\begin{center}
\includegraphics[width=\linewidth]{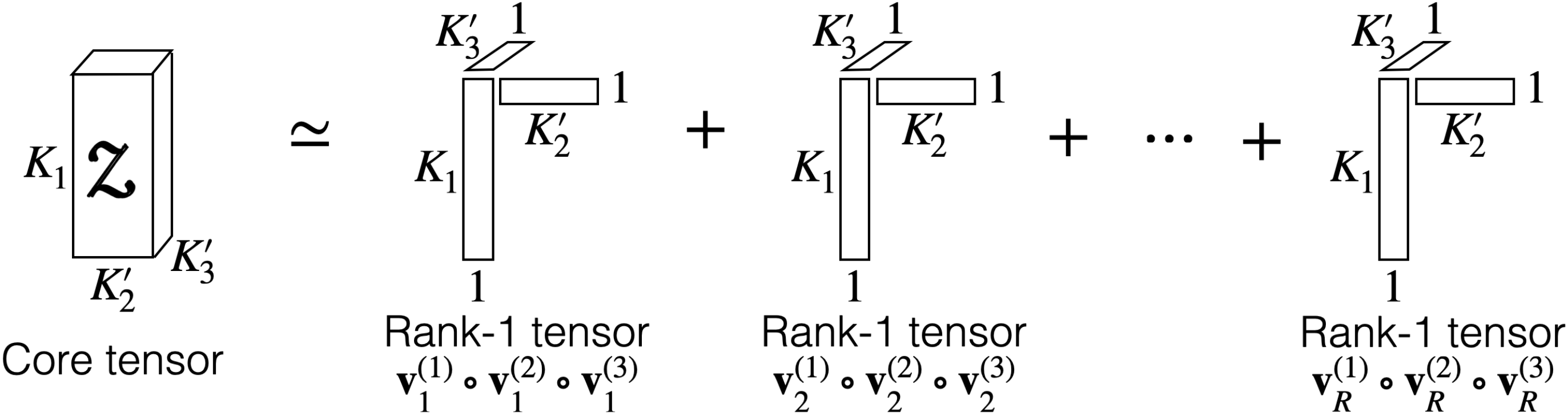}
\caption{The CP decomposition.}
\label{fig:CPDecomposition}
\end{center}
\end{figure}

\subsection{Core Tensor Visualization}
\label{sec:core-tensor-vis}

We introduce a core tensor visualization method that is designed to effectively summarize the core tensor with a small set of simple 2D visualizations.
To easily visualize an $\NModes$-mode core tensor, we first apply the CP decomposition to the core tensor.
The CP decomposition approximates the core tensor as a sum of $\CPDecompR{}$ rank-1 tensors for a user-specified $\CPDecompR{}$.
\autoref{fig:CPDecomposition} shows a graphical explanation of the CP decomposition.
More precisely, the CP decomposition of $\smash{\Tensor{Z}}$ is written as:
\begin{equation}
    \Tensor{Z} \simeq \sum_{\CPDecompIndex{}=1}^{\CPDecompR{}} \CPFactor{1}{\CPDecompIndex} \circ \CPFactor{2}{\CPDecompIndex} \circ \cdots \circ \CPFactor{\NModes}{\CPDecompIndex}
\end{equation}
where $\CPDecompIndex{}$ is the index of the rank-1 tensors, $\smash{\CPFactor{\Mode}{\CPDecompIndex}} \in \mathbb{R}^{\NFibers{\Mode}'}$ is a vector corresponding to mode-$\Mode$ for the $\CPDecompIndex{}$-th rank-1 tensor, and $\circ$ represents the vector outer product.
For each mode, we can collect a set of the vectors composing the rank-1 tensors, i.e., \scalebox{0.9}{$\smash{\CPFactorMat{\Mode} = (\CPFactor{\Mode}{1} \, \cdots \, \CPFactor{\Mode}{\CPDecompR{}})}$}. This set is called a \textit{factor matrix} for mode-$\Mode$.
The resultant factor matrices summarize the essential information of the core tensor, $\Tensor{Z}$.

For mode-$1$ (i.e., the mode where groups are defined), we visualize $\smash{\CPFactorMat{1}}$ as a 2D scatterplot by setting $\CPDecompR{} = 2$ by default (\autoref{fig:teaser}-b) for two reasons.
First, a 2D scatterplot is commonly used to visualize DR results~\cite{Fujiwara:2022:ULCA,mcinnes2018umap}, making it a familiar visualization for analysts. 
Second, a 2D scatterplot can clearly depict distance relationships among points, highlighting proximity-related patterns, such as clusters and outliers.
For example, in \autoref{fig:teaser}-b, we can see Groups 1--3 have clear separation while Group 2 and Group 3 form subclusters.

For the other modes, we want to convey how these modes are associated with each axis of the 2D scatterplot (i.e., \scalebox{0.9}{$\smash{\CPFactor{1}{1}}$} and \scalebox{0.9}{$\smash{\CPFactor{1}{2}}$}).
We visualize \scalebox{0.9}{$\smash{\CPFactor{\Mode}{1}}$} and \scalebox{0.9}{$\smash{\CPFactor{\Mode}{2}}$} ($\smash{\Mode \in \{2, \cdots, \NModes\}}$) using a set of bar charts.
\autoref{fig:teaser}-c shows an example of the bar charts, which are arranged in order of \scalebox{0.8}{$\smash{\CPFactor{2}{1}}$}, \scalebox{0.8}{$\smash{\CPFactor{2}{2}}$}, \scalebox{0.8}{$\smash{\CPFactor{3}{1}}$}, \scalebox{0.8}{$\smash{\CPFactor{3}{2}}$} (note: mode-2 is the instance mode and mode-3 is the variable mode in this example).
From \autoref{fig:teaser}-c, we can infer that the $y$-axis of the scatterplot (i.e., \scalebox{0.85}{$\smash{\CPFactor{1}{2}}$}) is highly associated with the first instance component and the second variable component.

Thus, we can visualize an $\NModes$-mode core tensor with one 2D scatterplot and $2(\NModes - 1)$ bar charts.
We assume $\NModes$ and, as a by-product, the required visualizations are also relatively small  (e.g., $\NModes = 3$ or 4).

\subsection{Visual Analytics Interface}
\label{sec:system}

We design a visual analytics interface to effectively visualize all the essential information of TULCA (\autoref{sec:essential-info}). 
The implemented interface is shown in \autoref{fig:teaser}.
This interface is a prototype specific for a third-order tensor with time, instance, and variable modes. 
This interface can easily be modified or extended to support higher-order tensors (e.g., add more bar charts in \autoref{fig:teaser}-c based on the additional number of modes).

\textbf{TULCA parameters.}
Analysts can flexibly control TULCA using the key parameters listed in \autoref{fig:teaser}-a.
First, they select the mode to treat as mode-$1$ (i.e., the mode with groups that should be compared to) from ``Comparing mode''.
Second, they specify the number of components TULCA extracts from each mode (i.e., $\NFibers{\Mode}'$) using the ``\# of components'' dropdowns. 
Third, they set the weight parameters (``Target Weight'', ``Background Weight'', ``Between-class Weight'') based on their analytical goal.
These weights are shown as interactive bar charts, adjustable via draggable tabs. 
Analysts can also modify weights for multiple groups at once by checking boxes next to group names.
Any change to these key parameters automatically updates TULCA results and related views.

\textbf{Core tensor.}
We visualize the core tensor using a 2D scatterplot and a set of vertical bar charts (\autoref{fig:teaser}-b, c) (cf.  \autoref{sec:core-tensor-vis}).
Analysts can select a subset of visualized points (i.e., timepoints in \autoref{fig:teaser}) in the scatterplot.
Despite the subset selection, analysts can always refer to the original context of the data with \autoref{fig:teaser}-f. 
We discuss more below.

\begin{figure}[t]
    \centering
    \includegraphics[width=\linewidth]{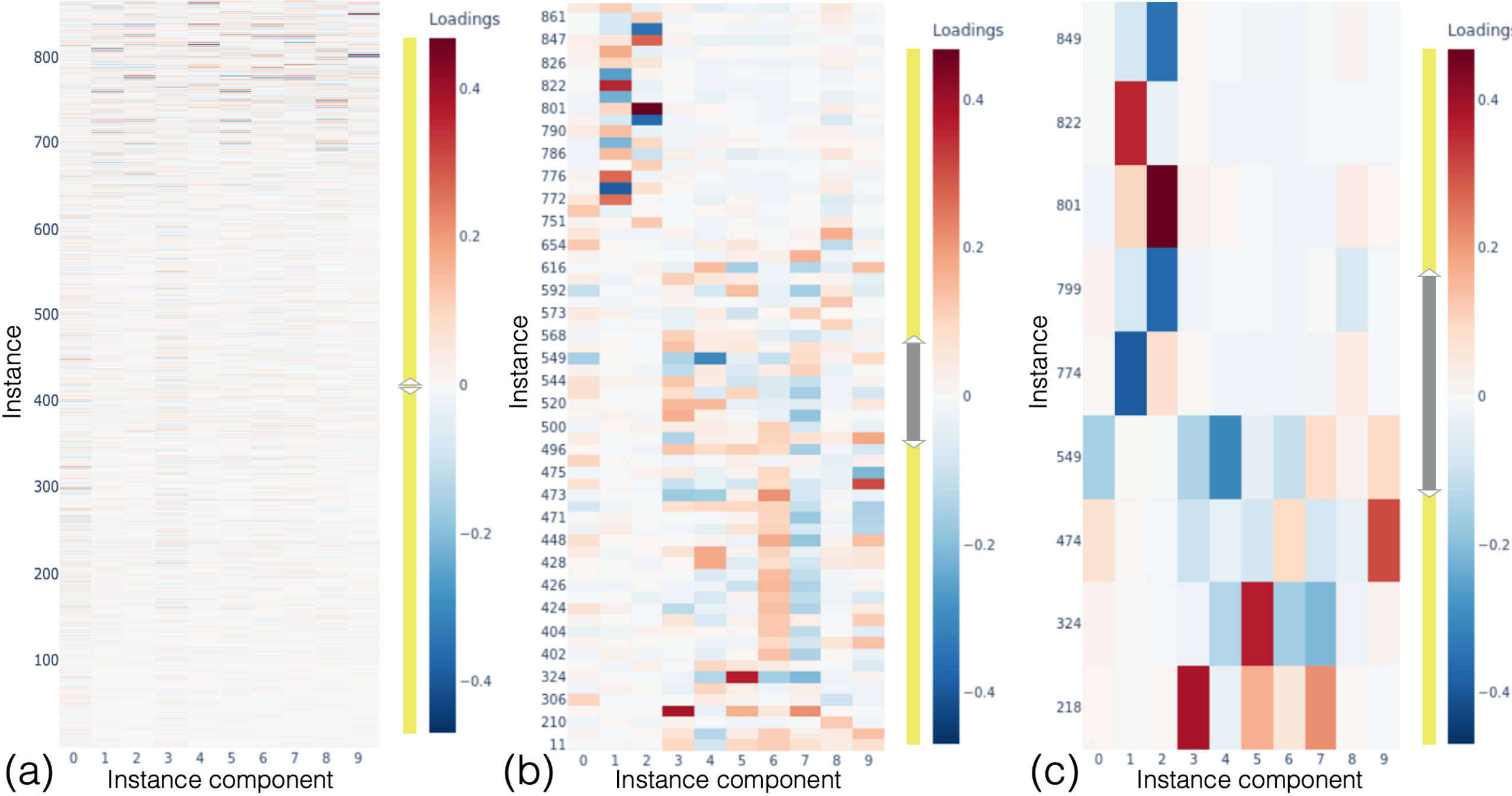}
    \caption{Filtering of projection matrix values using a colormap range slider: (a) the original matrix with 864 rows/instances; (b, c) the matrices after filtering, where 64 and 9 rows are visualized, respectively.}
    \label{fig:filtering}
\end{figure}

\textbf{Projection matrices.}
The projection matrices ($\ProjMat{2}, \cdots, \ProjMat{\NModes}$) define a multilinear mapping from the original dimensions to the components.
Their values, called \textit{loadings} are visualized as blue--red divergent color heatmaps with rows as the original dimensions and columns as the components (\autoref{fig:teaser}-d, e).
A darker color indicates the corresponding original dimension more strongly influences the component.
Hues indicate the direction of their influence.
As shown in \autoref{fig:filtering}-a, when many dimensions exist (e.g., 864 rows), the heatmap assigns a limited height to each row, reducing visibility.
To address this visual scalability issue, the interface provides a colormap range slider on the left side of each colormap.
\autoref{fig:filtering}-b, c demonstrate this functionality, where 64 and 9 out of 864 rows are visualized, respecively. 
Adjusting the slider will filter \textit{out} dimensions whose loadings fall entirely within the selected range (colored gray in \autoref{fig:filtering}-b, c).
This filtering approach is motivated by the importance of finding dimensions with significant influence on the projection over those with negligible impact.

By concurrently checking these projection matrices and the bar charts in \autoref{fig:teaser}-c, we can understand how the $x$- and $y$-axes in the scatterplot relate to the original dimensions.
From \autoref{fig:teaser}-c, we can see that the second variable component is heavily associated with both $x$- and $y$-axes of the scatterplot. 
We showcase simple examples in \autoref{sec:synthetic}.

\textbf{Original data.}
In \autoref{fig:teaser}-f, we visualize the original data corresponding to the selection made in the scatterplot (\autoref{fig:teaser}-b).
How the original data should be visualized is highly dependent on the analysis targets and domain applications.
Example plots shown in \autoref{fig:teaser}-f are designed for analyzing supercomputer log data.
We demonstrate these plots usage in \autoref{sec:case-supercomputer}.
Designing a universal solution that fits any domain application is infeasible. We instead designed the interface (\autoref{fig:teaser}a--f) to be seamlessly integrated with Python and the Jupyter Notebook to  enable customizing visualizations based on analytical needs.

\textbf{Implementation.}
The interface is implemented with Python3 and Plotly Dash~\cite{dash}.
This implementation enables analysts to use our visualizations in a Jupyter Notebook environment while using external libraries in conjunction with our interface.
For the CP decomposition, we used TensorLy's implementation with default parameters (e.g., using the singular value decomposition to initialize factor matrices~\cite{kolda2009tensor}).

\section{Experiment Using Synthetic Data}
\label{sec:synthetic}

\begin{figure}[t]
\begin{center}
\includegraphics[width=\linewidth]{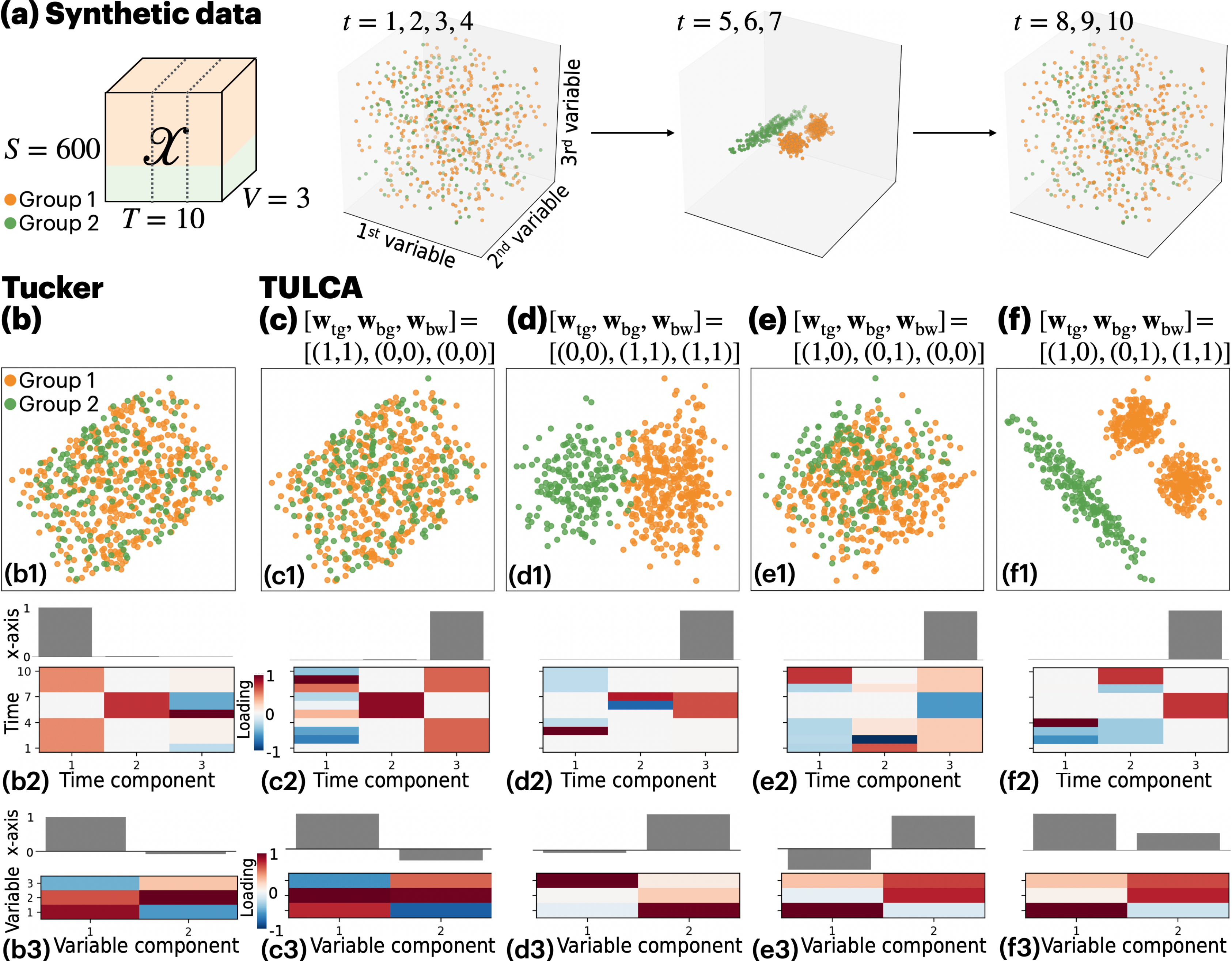}
\caption{Comparison of tensor decomposition results using synthetic data: (a) the synthetic data, (b) Tucker decomposition results, and (c--f) TULCA results using various parameters. (b1--b3) show the plots employed by the visual analytics interface: (b1) the 2D scatterplot of the core tensor, (b2) the time component information, and (b3) the variable component information (bar chart: the rank-1 tensor obtained by the CP decomposition, heatmap: the projection matrix). (c1--c3), (d1--d3), (e1--e3), and (f1--f3) present the same set of plots as (b1--b3).}
\label{fig:synthetic}
\end{center}
\end{figure}

To test TULCA's functionality, as shown in \autoref{fig:synthetic}-a, we designed synthetic multivariate time-series data consisting of 600 instances, 3 variables, and 10 timepoints (i.e., 3D data evolving over time)~\cite{supp}.
We assigned 400 instances to Group 1 (orange) and 200 to Group 2 (green).
All instances were initialized with random 3D positions, which remained unchanged at $t {=} 1, 2, 3, 4$ and $t {=} 8, 9, 10$.
For $t {=} 5, 6, 7$, we generated two isotropic Gaussian blobs for Group 1 and one anisotropic blob for Group 2.
This setup simulates a realistic scenario where only a few timepoints reveal distinct group patterns and some group contains hidden subgroups (Group1 has two subgroups in the synthetic data).

As shown in \autoref{fig:synthetic}, we applied (b) Tucker decomposition as a baseline and (c--f) TULCA with different parameters to the synthetic data.
We set the core tensor size to $S {\times} V' {\times} T' {=}\, 600 {\times} 2 {\times} 3$ for all results. 
We can observe different patterns from the 2D scatterplots of the extracted core tensors (b1--f1). 
Tucker decomposition, which aims to preserve overall variance, tends to capture random noise in the data.
TULCA with $\smash{\Mat{\TargetWeight{}} {=} (1, 1), \Mat{\BackgroundWeight{}} {=} (0, 0), \Mat{\BetweenWeight{}} {=} (0, 0) }$ focuses on preserving all groups' variances (\autoref{fig:synthetic}-c1), producing similar results to Tucker decomposition.
In contrast, using $\smash{\Mat{\TargetWeight{}} {=} (0, 0), \Mat{\BackgroundWeight{}} {=} (1, 1), \Mat{\BetweenWeight{}} {=} (1, 1)}$ (equivalent to TDA) results in a clear separation between Groups 1 and 2 (\autoref{fig:synthetic}-d1); however, since TDA minimizes within-group variances, it fails to reveal Group 1’s subgroups.
With $\smash{\Mat{\TargetWeight{}} {=} (1, 0), \Mat{\BackgroundWeight{}} {=} (0, 1), \Mat{\BetweenWeight{}} {=} (0, 0)}$ (i.e., TcPCA), TULCA shows higher variance in Group 1 but does not reveal its subgroups (\autoref{fig:synthetic}-e1) likely due to large impact from the random noise at $t {=} 1, 2, 3, 4$ and $t {=} 8, 9, 10$.
Finally, with $\smash{\Mat{\TargetWeight{}} {=} (1, 0), \Mat{\BackgroundWeight{}} {=} (0, 1), \Mat{\BetweenWeight{}} {=} (1, 1)}$, TULCA shows both the separation between Groups 1 and 2 and the subgroups within Group 1 (\autoref{fig:synthetic}-f1) by combining discriminant analysis and contrastive learning.

We can further interpret how each method generated the scatterplot result by reviewing the other plots in the visual analytics interface. 
In \autoref{fig:synthetic}-b2--f2, the bar charts show the time components' associations with the scatterplots' $x$-axes while the heatmaps show how the time components are derived from the original timepoints.
\autoref{fig:synthetic}-b3--f3 show the same sets of visualizations for the variable components.
We can see that the Tucker decomposition result mainly reflects the first time component (see the bar chart in b2), which is associated with $t=1, 2, 3, 4$ and $t=8, 9, 10$ (the heatmap in b2).
In contrast, the TDA result focuses on the third time component corresponding to $t=5, 6, 7$ (\autoref{fig:synthetic}-d2), leading to the group separation.
However, the TDA result relies on the second variable component (the bar chart in d3), mainly derived from the first variable (the heatmap in d3), and does not find the subgroups of Group 1.
Lastly, from \autoref{fig:synthetic}-f2--3, the TULCA result in (f) mainly captures the patterns in $t=5, 6, 7$ (similar to TDA) while utilizing the first and second variable components that cover all three variables, resulting in the pattern seen in \autoref{fig:synthetic}-f1.

The above results demonstrate the effectiveness of TULCA’s flexible tensor group comparison in uncovering various hidden patterns.

\vspace{5pt}
\section{Case Studies}
\label{sec:case-studies}

To investigate the effectiveness of the TULCA-based visual analytics, we conducted two case studies using real-world multivariate time-series data: supercomputer and mobile health~\cite{adibi2015mobile} log data. An operational staff member from the supercomputer center (RIKEN R-CCS), representing a target user of our visual analytics system, participated in the case studies.
The operational staff served as a domain expert for Case Study 1 and as a non-domain expert for Case Study 2.

\subsection{Case Study 1: Supercomputer Log Data}
\label{sec:case-supercomputer}

We analyze the operational log data collected from the K computer~\cite{miyazaki2012overview}, which includes four sensor-based temperature measurements from the compute racks: cooling air intake temperature (\texttt{AirIn}); cooling air exhaust temperature (\texttt{AirOut}); average CPU temperature (\texttt{CPU}); and cooling water inlet temperature (\texttt{Water}). 
The K computer consists of 1,080 racks: 864 compute racks and 216 disk racks.
These racks are spatially ordered from the ground up, indexed from 0 to 44 in the X-axis direction and from 0 to 23 in the Y-axis direction. 
In this case study, we are interested in analyzing the operational logs related to only compute racks.
The K computer's log data was collected during the Japanese fiscal year (FY) (i.e., starting in April and ending in March).
We analyze the daily average log data collected from FY2014 to FY2016, resulting in a third-core tensor with $\NTimes{=}1086$, $\NInsts{=}864$, and $\NVars{=}4$ (in total: 3,753,216 elements).

\begin{figure}[t]
\begin{center}
\includegraphics[width=\linewidth]{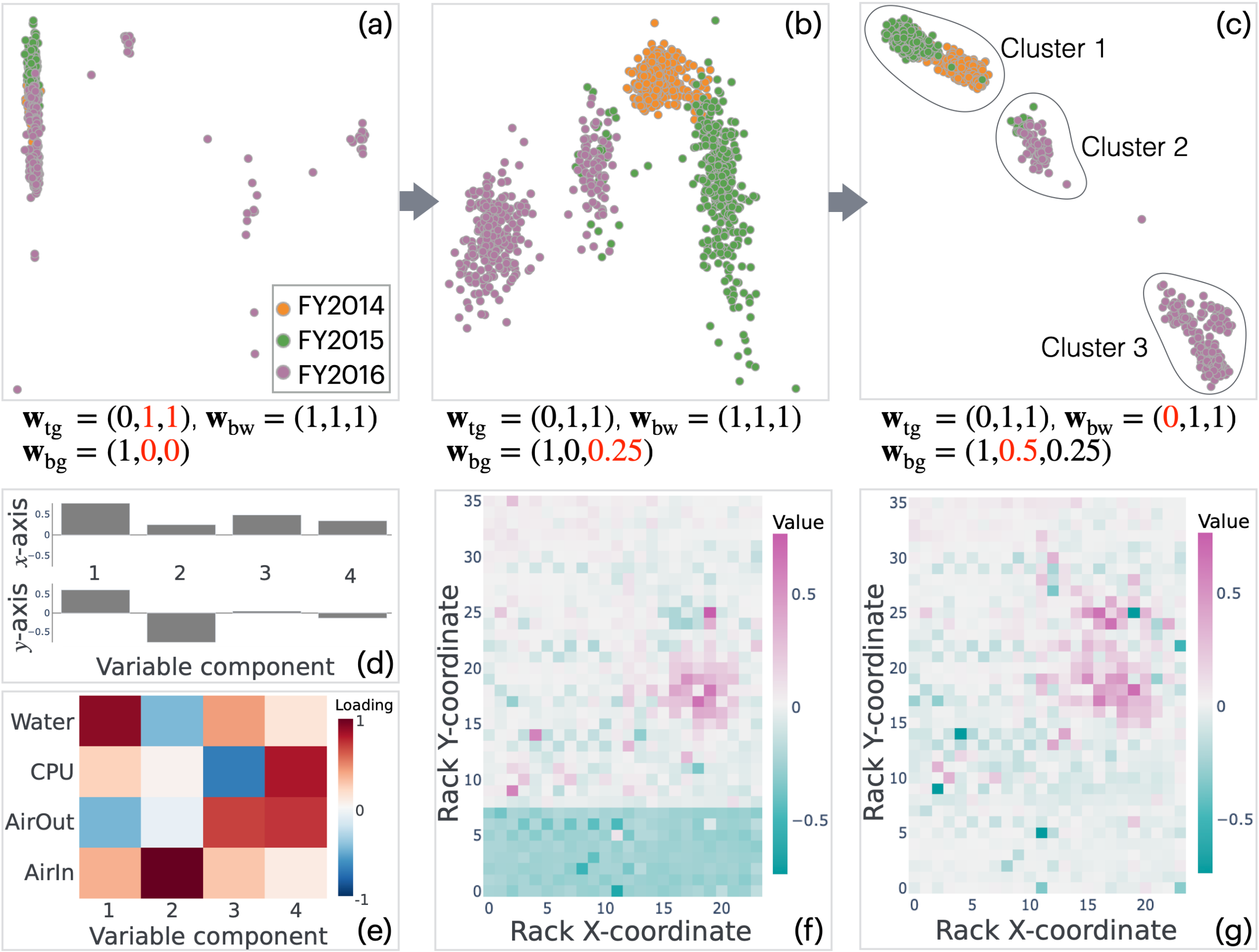}
\caption{Case Study 1: Identifying unique patterns in each fiscal year's supercomputer operations of the water/air cooling infrastructure}.
\label{fig:case1}
\end{center}
\end{figure}

In this analysis, we aim to uncover patterns unique to each fiscal year. 
Although the K computer itself was already decommissioned, the water/air cooling facility has been used for the current supercomputer (Fugaku), and it is planned to be operational until the beginning of the operation of the next supercomputer scheduled for 2030. 
Therefore, a better understanding of the cooling facility operation is demanded for the current operational planning and execution improvement. For this purpose, we first apply the weight parameters that are equivalent to TDA (i.e., \autoref{fig:teaser}-a).
The result shows a clear separation of each fiscal year, suggesting each year has some operational differences.
We then want to focus on variance patterns that can be only seen in more recent years (i.e., \texttt{FY2015} and \texttt{FY2016}) while maintaining separation between each fiscal year.
To find such patterns, we update the parameters in \autoref{fig:case1}-a (updated parameters are colored red) to maximize the variance preservation of \texttt{FY2015} and \texttt{FY2016} while reducing the variance of \texttt{FY2014}.
The output result is heavily influenced by the variance of \texttt{FY2016} and largely highlights various outliers from \texttt{FY2016}. 
Though investigating these outliers is useful to find anomalies within a short period (e.g., a day), we rather focus on finding patterns related to a longer period.
We reduce the variance of \texttt{FY2016} to eliminate \texttt{FY2016}'s outliers by increasing the background weight for \texttt{FY2016}, as shown in \autoref{fig:case1}-b.
In \autoref{fig:case1}-c, we further adjust the parameters to provide more separation between each visually-identified cluster, resulting in three clusters: Clusters 1--3.
While \texttt{FY2014} and \texttt{FY2015} form Cluster 1, \texttt{FY2016} is separated into Clusters 2 and 3. 
According to the operational staff, the utilization rate at the beginning of each fiscal year has been around 50--70\%. However, in \texttt{FY2016}, it increased to 70--80\% due to the introduction of low-priority jobs, which may explain the separation into two clusters.
We now review the clusters more closely.

\autoref{fig:case1}-d shows the variable components' associations with $x$- and $y$-axes of \autoref{fig:case1}-c.
This results shows that the first variable component has high associations for both axes.
By checking the variable mode's projection matrix (\autoref{fig:case1}-e), we confirm the first variable component is mainly derived from \texttt{Water}.
Thus, we decide to examine the detailed differences in \texttt{Water} between the three identified clusters (\mbox{Clusters 1--3}).
We first select Cluster 1 and Cluster 2 and review their differences in  \texttt{Water} (\autoref{fig:case1}-f).
\autoref{fig:case1}-f shows subtraction difference between \texttt{Water}'s mean value of the timepoints in Cluster 1 and the mean value of Cluster 2's timepoints, i.e., (Cluster 2's mean) $-$ (Cluster 1's mean).
We clearly observe that compute racks with Y-coordinates 0--7 have negative values, indicating \texttt{FY2016} in Cluster 2 has a lower \texttt{Water} temperature than \texttt{FY2014} and \texttt{FY2015}. 

According to the operational staff, this finding corresponds to the gradual modifications to the cooling infrastructure throughout \texttt{FY2016}.
We have also identified that some compute racks in Cluster 3 have significantly lower temperatures than those in Cluster 1. The operational staff verified this fact and confirmed that these compute racks temporarily suffered from measurement failures. 

\textbf{Expert feedback.}
The above results highlight the effectiveness of TULCA in finding operational peculiarities and anomalies which was not possible with the former DR-based visual analytics system that the operational staff was accustomed to working with. 
The operational staff was able to easily replicate these results by installing the prototype system in his daily working environment.
After interacting with the system, he expressed its high usefulness, including TULCA's flexibility for grouping or separating clusters and the interactive comparison of two selected clusters via the original data plots (\autoref{fig:teaser}-f). 
For instance, he informed that it was very meaningful to be able to identify subclusters corresponding to a specific period, such as the beginning or end of the fiscal year. 
Additionally, it was also intuitive for him to see the differences between clusters directly in a heatmap with the physical locations of the computational racks.
He also positively evaluated the interactive functionalities, such as filtering and brushing, described in \autoref{sec:system}. 
However, he also noted concerns rooted in the lack of technical knowledge of tensor decomposition and TULCA. 
He was not able to effectively utilize the other visualizations, such as the projection matrix heatmaps and core-tensor bar charts (\autoref{fig:teaser}-c, d, e).
But, this feedback also implies the original data plots (e.g., \autoref{fig:teaser}-f) are highly helpful to improve the interpretability of tensor decomposition results.
Also, adjusting the weight parameter was mentally demanding for him since it sometimes caused significant visual changes in the scatterplot of the TULCA result.
We discuss potential improvements on the parameter selection in \autoref{sec:discussion}.

\vspace{-3pt}
\subsection{Case Study 2: Mobile Health Log Data}
\label{sec:mhelath}
\vspace{-2pt}

We analyze a publicly available mobile health dataset  (MHEALTH)~\cite{banos2014villalonga,banos2015mhealth}. This dataset includes 23 sensor motion measures collected by 10 subjects where sensors were placed on each subject's right wrist, left ankle, and chest. The sensors recorded acceleration, magnetic field, and rotational speed in each direction (X, Y, Z) at a sampling rate of 50 Hz. 
Each subject performed 12 pre-defined physical activities. 
In this case study, we compare 8 physical activities performed by the 10 subjects over 1-minute period: (1) \texttt{standing still}, (2) \texttt{sitting and relaxing}, (3) \texttt{lying down}, (4) \texttt{walking}, (5) \texttt{climbing stairs}, (6) \texttt{cycling}, (7) \texttt{jogging}, (8) \texttt{running}.
Though these activities are measured at a high sampling rate (i.e., 50 Hz or every 0.02 second), we reduce the data size by extracting each measure for a 1-second interval.
The resulting tensor has 480 timepoints (60 seconds\,$\times$\,8 activities), 10 subjects/instances, and 23 measures/variables.

\begin{figure}[t]
\begin{center}
\includegraphics[width=\linewidth]{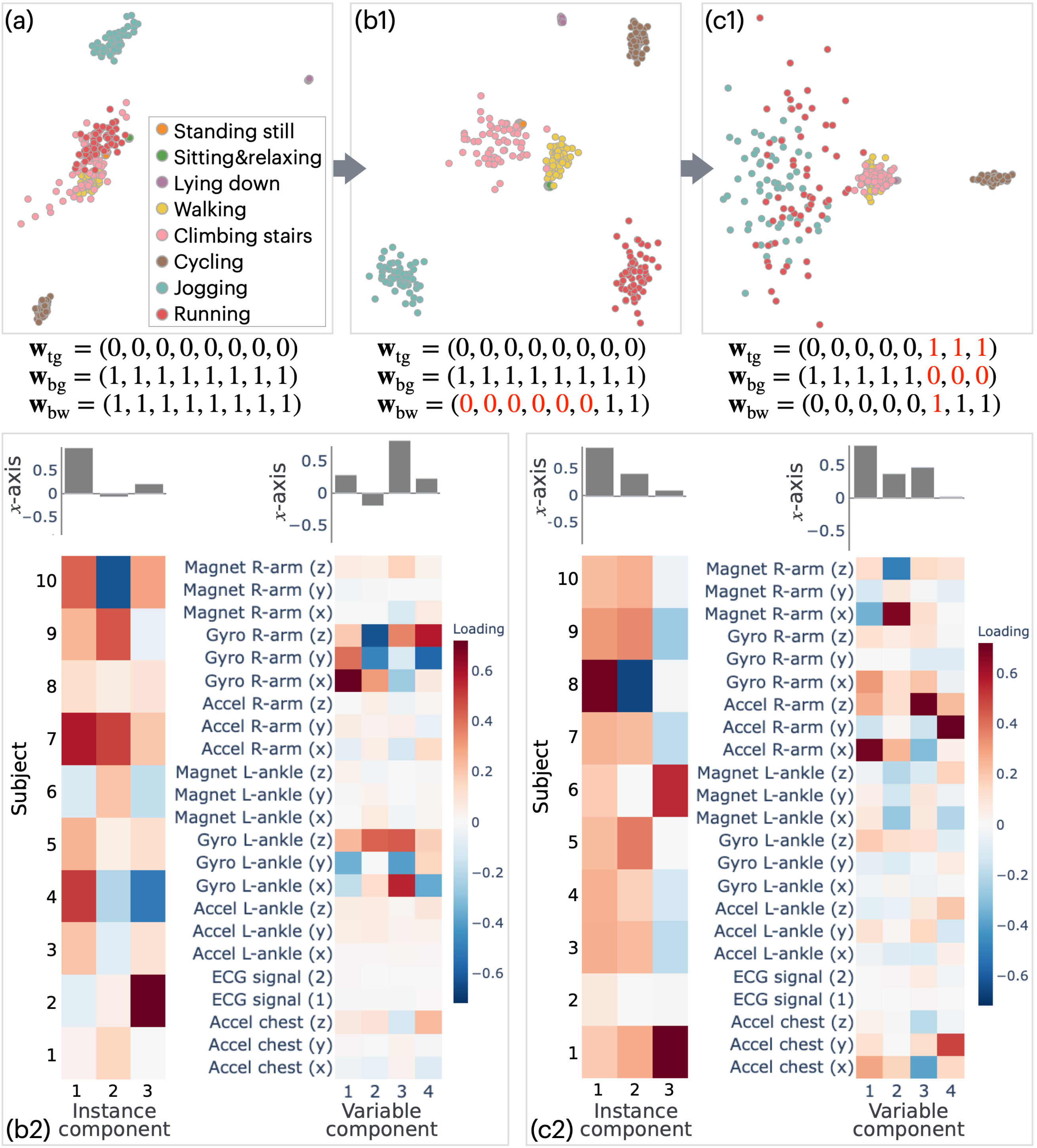}
\caption{Case Study 2: Investigating unique patterns in Jogging and Running from the MHEALTH log data.}
\label{fig:case2}
\end{center}
\end{figure}

We first apply the weight parameters equivalent to TDA, producing the result shown in \autoref{fig:case2}-a, where the scatterplot points correspond to timepoints.
We can see \texttt{lying down} (purple points) is spatially separate from the other activities.
This result only leads to a trivial insight (i.e., \texttt{lying down} is clearly different from the other activities). 
As such, emphasizing the difference of \texttt{lying down} in the core tensor is likely degrading the separation among the other activities. 

Subsequently, we shift our focus to analyze two similar activities, \texttt{jogging} (teal) and \texttt{running} (red) to understand their differences and similarities.
To find the differences, we set large between-class weights only for these two activities (\autoref{fig:case2}-b1). 
The result shows a clear separation along the $x$-axis.
\autoref{fig:case2}-b2 shows the projection matrices used to generate \autoref{fig:case2}-b1.
At the top of each projection matrix, the height of each bar charts shows each component's association contributing to the scatterplot's $x$-axis.
These bar charts identifies that the first instance component and the third variable component contribute most to the \textit{difference} between \texttt{jogging} and \texttt{running}.  
The first instance component of the projection matrix shows all subjects except for subjects 2 and 6 have positive loadings.
This pattern indicates that the other 8 subjects have a similarity in the differences of their \texttt{jogging} and \texttt{running} movements.
In addition, from the third variable component in the projection matrix, we can see that \texttt{Gyro L-ankle (x)}, \texttt{Gyro L-ankle (y)}, \texttt{Gyro L-ankle (z)}, and \texttt{Gyro R-arm (z)} are strongly related to the separation of \texttt{jogging} and \texttt{running}. 
These variable loadings show that the arm's rotation speed in the Z-direction and ankle's rotation speed in all X-, Y-, Z-directions are clearly different when \texttt{jogging} versus \texttt{running}. 
Furthermore, \texttt{cycling} (brown) shares similar $x$-coordinates with \texttt{running}, indicating that \texttt{cycling} has similar movement patterns as \texttt{running} for these measurements.

Based on this high-level insight on how \texttt{cycling} and \texttt{running} share similarities, we are now interested in to identifying patterns related to the three activities (\texttt{cycling}, \texttt{running}, and \texttt{jogging}) compared to the other five activities.
Various patterns can be revealed by adjusting the weight parameter setting, and here, we demonstrate one such example.
In \autoref{fig:case2}-c1, we increase the between-weight parameter for \texttt{cycling} to increase separation among the three activities,  increase target weights for \texttt{cycling}, and reduce the background weights of the other five activities to further contrast their variances. 
The $x$-axis in \autoref{fig:case2}-c1 shows how \texttt{cycling} is clearly separated from \texttt{running} and \texttt{jogging}. 
However, \texttt{running} and \texttt{jogging} have significant overlaps while also being widely scattered compared to the other five activities.
The wide-spread distribution of \texttt{running} and \texttt{jogging} indicates large variances in the two activites.
By looking at the projection matrices in \autoref{fig:case2}-c2, we observe that the first instance component is highly associated with the $x$-axis of the scatterplot and has positive loadings for all subjects (i.e., all subjects share the similarities for this factor).
The variable projection matrix also shows that the first variable component is most associated with the $x$-axis and has a distinct and large loading for \texttt{Accel R-arm (x)} (i.e., the acceleration of the right arm for $x$-direction). 
This result matches with our general intuition: we introduce acceleration through our arms when \texttt{running} or \texttt{jogging} but less so when \texttt{cycling} (i.e., arms are generally stationary). 
This results also highlights how the degree of the acceleration differs person by person or even within a movement as quickly as seconds.

This case study demonstrates TULCA's ability to identify three different patterns using three comparative analysis methods: discriminant analysis (\autoref{fig:case2}-a), enhancement of discriminant analysis (\autoref{fig:case2}-b1), and simultaneous combination of discriminant analysis and contrastive learning (\autoref{fig:case2}-c1). Extracting all three patterns is possible given how we can flexibly adjust TULCA's weight parameters.
The supercomputer operational staff also replicated these results in his working environment and expressed the system’s usefulness based on their own analysis. However, he noted that adjusting multiple weight parameters individually was labor-intensive, prompting the development of grouped parameter adjustment as described in \autoref{sec:system}.

\begin{table}[t]
  \caption{Completion time in seconds. Values under each dataset name indicate their tensor size (i.e., $\NFibers{1}{\times}\NFibers{2}{\times}\NFibers{3}$).}
  \label{table:time}
  \renewcommand{\arraystretch}{0.8}
  \scriptsize
  \setlength{\tabcolsep}{0.2em}
  \centering
  \begin{tabular}{lrrrr}
        & \textbf{US Air} & \textbf{High School} & \textbf{MHEALTH} & \textbf{K Log} \\
        & $53{\times}55{\times}5$ & $180\times{30}\times{9}$ & $480{\times}10{\times}23$ & $1086{\times}864{\times}4$ \\
    \toprule 
    \texttt{Tucker} & 0.005 & 0.009 & 0.018 & 3.849 \\
    \texttt{TULCA-all} & 0.008 & 0.006 & 0.005 & 11.389 \\
    \texttt{TULCA-update} & 0.001 & 0.001 & 0.001 & 0.573 \\
    \bottomrule 
  \end{tabular}
\end{table}

\begin{figure}[t]
\begin{center}
\includegraphics[width=\linewidth]{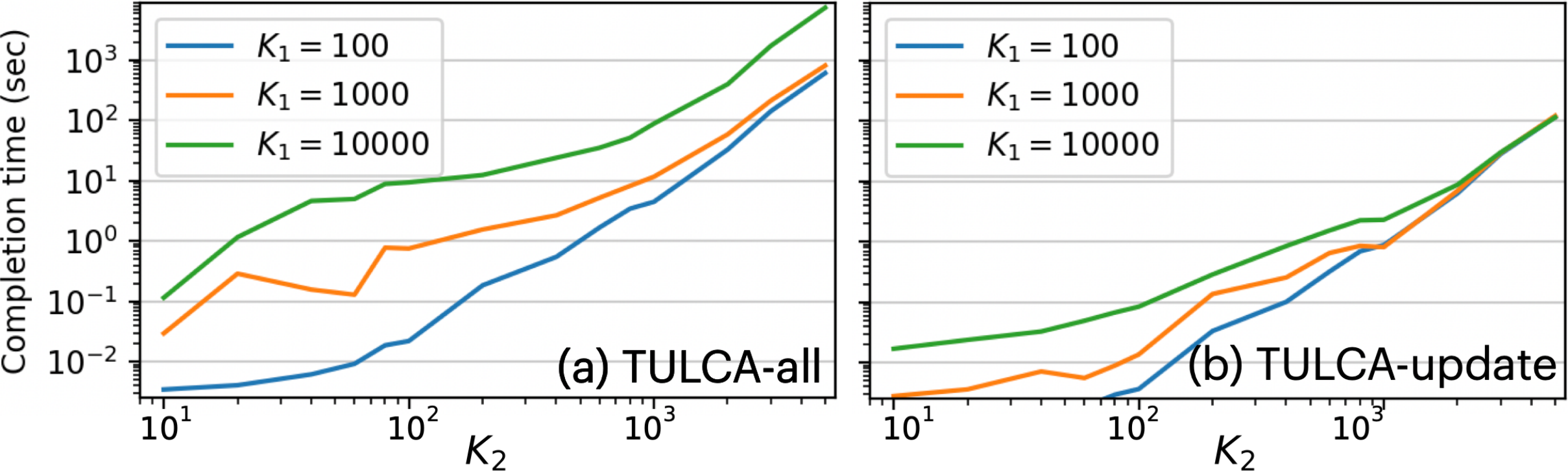}
\caption{Completion time of TULCA for different sizes of tensors.}
\label{fig:perf}
\end{center}
\end{figure}

\section{Performance Evaluation}
\label{sec:performance}
We conducted a performance evaluation of TULCA by using a MacBook Pro (16-inch, 2019) with 2.3 GHz 8-Core Intel Core i9 and 64 GB 2,667 MHz DDR4. 
We measured TULCA's completion time for multiple real-world datasets: the K computer log data (\autoref{sec:case-supercomputer}), the MHEALTH dataset (\autoref{sec:mhelath}), the US air quality data~\cite{airdata}, and the high school dynamic contact networks~\cite{fournet2014contact}. \autoref{table:time} summarizes the tensor size of each data.
We used versions of the US air quality data and the high school dynamic contact networks that are processed as third-order tensors~\cite{Fujiwara:2021:MultiDR}.
We evaluated three measures of completion time: (1) TULCA's entire process (\texttt{TULCA-all}), which also includes its covariance matrix computations, (2) TULCA's update time based on weight parameter changes (\texttt{TULCA-update}), and (3) \texttt{Tucker decomposition} as a representative baseline tensor decomposition method. 
We reduced the lengths of each dataset's mode-2 and mode-3 to $3{\times}3$ (e.g., extracting a $480{\times}3{\times}3$ core tensor for the MHEALTH dataset).
We executed each of the three measures 10 times and computed the average completion time. 
For each execution, we set TULCA's weight parameters to random values in $[0, 1]$.
\autoref{table:time} shows the evaluation result. 

\autoref{table:time} shows that TULCA is generally has a similar completion time to Tucker decomposition.
For the K computer log data (the largest data), \texttt{TUCLA-all} had the longest completion time, which was over 11 seconds.
Despite this large dataset, \texttt{TULCA-update} took less than a second.
This result highlights that TULCA is efficient in supporting interactive analysis even for a considerably large dataset. 

We further generated various sizes of tensors from the MHEALTH dataset to identify the upper limit of TULCA can reasonably handle for interactive use.
Given how TULCA handles mode-$1$ and the other modes differently, we prepared datasets with $\NFibers{1} = \{100, 1000, 10000\}$ and $\NFibers{2} = [10, 5000]$ but with a fixed number of $\NFibers{3}$ ($\NFibers{3}=23$). 
We generated these datasets by sampling timepoints and repeating instances in the original MHEALTH data (e.g., applying a sampling rate of approximately 25 Hz to produce 10,000 timepoints, and repeating 10 original instances 10 times to create a total of 100 instances).
\autoref{fig:perf}-a, b show the completion times of \texttt{TULCA-all} and \texttt{TULCA-update}, respectively.
These results show that $\NFibers{2}$ has a stronger influence on the completion time for both \texttt{TUCLA-all} and \texttt{TULCA-update}.
When $\NFibers{1} = 10000$, $\NFibers{2} = 2000$, $\NFibers{3} = 23$ (460 million elements in total), \texttt{TULCA-all} completed in less than 7 minutes, while \texttt{TULCA-update} spent only 9 seconds.
In summary, TULCA is efficient in handling a large tensor (e.g., a half billion elements) for interactive analysis.

\section{Qualitative Comparison}

\begin{figure}[t]
\begin{center}
\includegraphics[width=\linewidth]{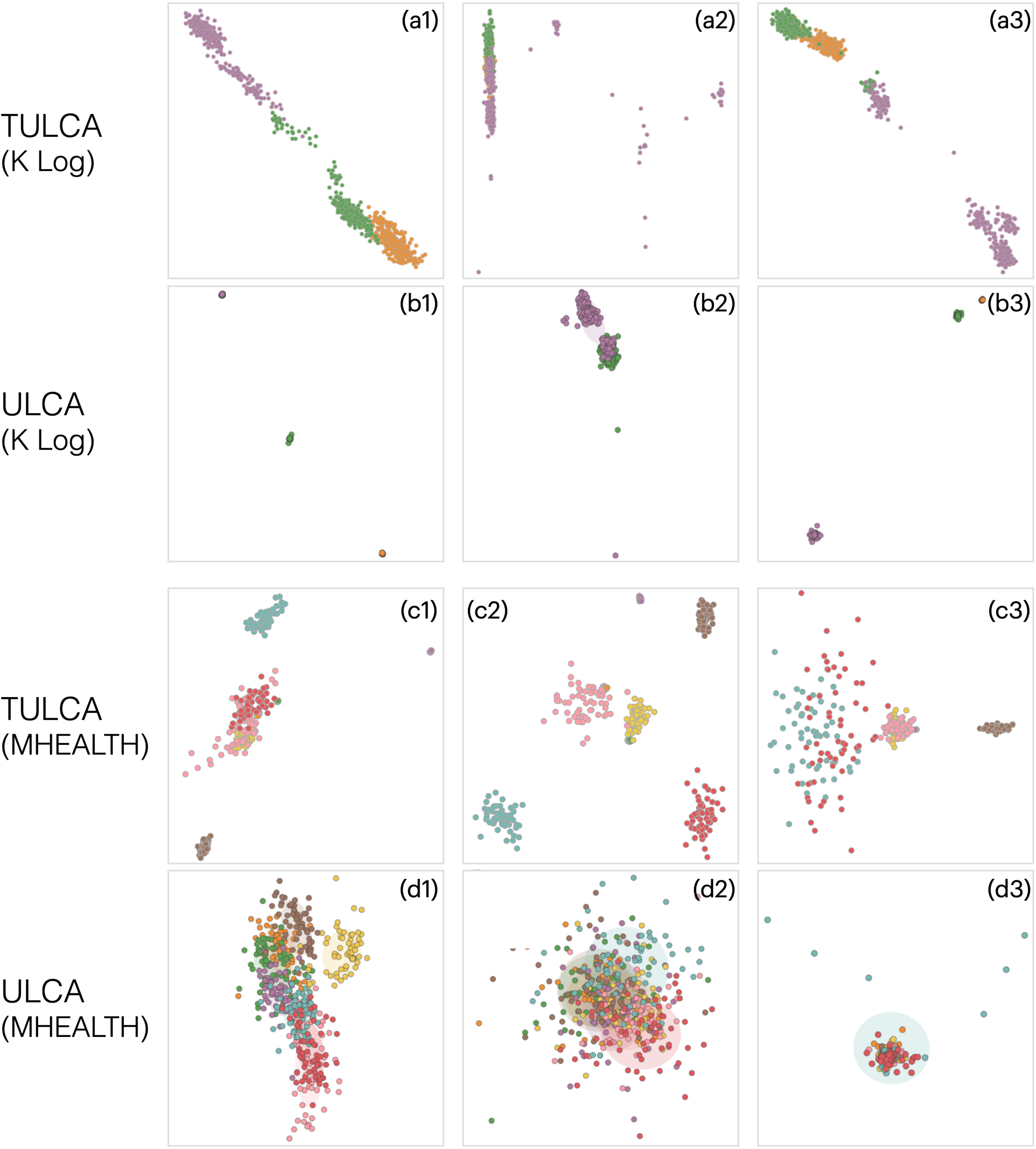}
\caption{Qualitative comparison of TULCA and ULCA results.}
\label{fig:tulca_vs_ulca}
\end{center}
\end{figure}

We quantitatively compare TULCA and ULCA results.
Though ULCA inherently focuses on second-order tensors, it can support higher-order tensors by first producing a mode-1 matricized tensor, $\Flatten{\Mat{X}}{1}$, and then apply ULCA to $\Flatten{\Mat{X}}{1}$.
As discussed in \autoref{sec:tensor-decomposition}, though matricizing a tensor and applying a single projection matrix summarizes the tensor, it often degrades the quality of the latent structure preservation~\cite{yang2004two,yang2005two}.
We expect such issues can also be seen in the ULCA results.
We apply ULCA to the K computer log and MHEALTH datasets using the same weight parameters we used to produce the TULCA results in \autoref{sec:case-studies}. 
The results are shown in \autoref{fig:tulca_vs_ulca}.

As seen in \autoref{fig:tulca_vs_ulca}-b1, b2, b3, ULCA tends to create very compact clusters compared to TULCA results. Consequently, it is difficult to extract interesting patterns as-is from ULCA results.
We expect that these compact clusters are due to the extremely high dimensionality relative to the number of points (specifically, 3456 dimensions vs. 1086 points). 
Extremely high dimensionality causes overfitting when using discriminant analysis. 
Given how a high-order tensor can easily induce this high dimensionality issue when matricized, these results highlight the critical limitation of ULCA when analyzing high-order tensors.

The MHEALTH data results (\autoref{fig:tulca_vs_ulca}-d1, d2, d3) do not show the same issue as the K computer log data, given that the MHEALTH dataset has only 480 points and 230 dimensions after mode-1 matricization.
However, for this dataset, ULCA does not show a clear separation among groups.
These results are consistent with Yang et al's findings~\cite{yang2004two,yang2005two}: the use of a tensor decomposition method, instead of a DR method, improves the quality of the latent structure preservation for post-analysis tasks (e.g., classification).
With ULCA, we cannot find various patterns and factors we identified in \autoref{sec:mhelath}, such as the factor differentiating \texttt{jogging} and \texttt{running}.

\section{Discussion}
\label{sec:discussion}

\textbf{Case Studies.}
Fiscal year comparisons from FY2014--FY2016 of the K supercomputer's log data in Case Study 1 highlight how TULCA can support analysts in investigating anomalies pertaining to one fiscal year compared to other years.
TULCA's efficacy stems from how analysts can easily adjust the weight parameters throughout their analytical workflow. 
For example, the weight parameters used in \autoref{fig:case1}-c enabled us to compare FY2014 /FY2015 and FY2016 to better understand the operation differences of the two fiscal years while still maintaining separation.
The MHEALTH dataset in Case Study 2 highlights TUCLA's ability to extract three different group comparisons: discriminant analysis (\autoref{fig:case2}-a), enhancement of discriminant analysis (\autoref{fig:case2}-b1), and simultaneous combination of discriminant analysis and contrastive learning (\autoref{fig:case2}-c1). These different patterns provide a deeper insight to understand what are the similarities and differences between different physical activities, such as \texttt{jogging}, \texttt{running}, and \texttt{cycling}.    
In \autoref{sec:mhelath}, we adjusted parameter weights to further explore the pattern of interest, highlighting TULCA's flexibility in supporting analysts to quickly pivot based on their analytical goal.

\textbf{Algorithm.}
Our work introduced TULCA as a primary step toward flexible comparative analysis of high-order tensors. 
However, TULCA needs further improvements to more effectively assist analysts, especially in its parameter selection.
TULCA encompasses both TDA (discriminant analysis) and TcPCA (contrastive learning), and combining both of them enables analysts to flexibly choose the comparative analysis method based on their analytical goal. 
However, from the case studies in \autoref{sec:case-studies}, we observed that trial-and-error is required to find parameters that produce interesting patterns. Therefore, future work should develop a parameter optimization method that finds the optimal weights as well as the optimal number of components ($\NFibers{\Mode}'$ for each mode-$\Mode$). 
The backward parameter selection in ULCA's implementation is one possible solution for the optimal weight setting. 
Fujiwara et al.~\cite{Fujiwara:2022:ULCA} showcase how this backward parameter selection can find ULCA's parameters that resemble user-demonstrated change over the visualized ULCA result (e.g., placing Group A close to Group B).
To select the optimal number of components, DIFFIT (the difference of fit)~\cite{timmerman2000three} is proposed for Tucker decomposition.  
Similar to the stopping criterion checking the learning curve in supervised learning, DIFFIT judges the appropriate number of components based on the improvement rate of data approximation when adding one extra component. 
As TULCA has a different optimization goal from Tucker decomposition, future research is required to find the best criterion to identify the optimal number of components for TULCA. 
Another potential issue is that flexible parameter selection may lead to spurious visual patterns (e.g., clusters, outliers), as observed in various DR methods~\cite{wang2021understanding,chari2023specious}.
Similar to DR~\cite{hyeon2025unveiling}, future research is needed to better understand the properties of TULCA and enhance the reliability of analyses using TULCA.
This includes sensitivity analysis of the weight parameters, as well as the development of quality metrics and comparison procedures for quantitatively evaluating interactive comparative analysis.

Our performance evaluation (\autoref{sec:performance}) demonstrated the computational efficiency of TULCA. 
However, further improvements will be necessary to scale to extremely large tensors.
One promising direction is the use of parallel and distributed computations.
For instance, the covariance matrix calculation could be computed using multiprocessing. 
As shown in \autoref{eq:cov-within-nmode}, the covariance matrix can be computed independently for each mode, slice, and group. 
In addition, a new parallel algorithm could be designed for the iterative eigenvalue decomposition, analogous to parallelized approaches for Tucker decomposition~\cite{kaya2016high}.

\textbf{Visualization Design.}
Our two case studies show that analysts can easily adjust TULCA's various parameters for comparative analysis. 
However, we acknowledge that datasets with large numbers of groups and modes can become difficult to identify and interpret the (dis)similarities of groups from the core tensor. 
For example, the parameter view assigns a separate bar chart for each group and weight, resulting in limited scalability.
Also, as the number of modes increases, the interface needs to add the corresponding core-tensor bar chart (\autoref{fig:teaser}-c) and projection matrix (\autoref{fig:teaser}-d) visualizations.
Future work should develop more scalable visualization and interaction methods.

\section{Conclusion}

We introduce TULCA, a new tensor decomposition method, that unifies discriminant analysis and contrastive learning for comparative analysis of high-order tensors. Analysts can flexibly switch between different comparative analysis schemes (i.e., discriminant analysis or contrastive learning) or even a mixture of both schemes depending on their analytical needs. 
With this unification of both schemes, TULCA facilitates
group comparisons for high-order tensors that cannot be achieved when solely using only one of the schemes.

\acknowledgments{%
  This work has been supported in part by the Knut and Alice Wallenberg Foundation through Grant KAW 2019.0024.
  The work of Takanori Fujiwara was completed when he was with Link\"{o}ping University.
}


\bibliographystyle{abbrv-doi-hyperref}

\bibliography{reference}

\begin{thebibliography}{10}

\bibitem{supp}
The supplementary materials: The source code for {TULCA} and the evaluations, along with a demonstration video of the visual interface.
\newblock \url{https://github.com/vizlab-kobe/tulca}.

\bibitem{abid2018exploring}
A.~Abid, M.~J. Zhang, V.~K. Bagaria, and J.~Zou.
\newblock Exploring patterns enriched in a dataset with contrastive principal component analysis.
\newblock {\em Nat Commun}, 9(1):2134, 2018. \href{https://doi.org/10.1038/s41467-018-04608-8}
{doi: {{%
10\hspace{.1pt}\discretionary{.}{%
}{.}\hspace{.4pt}1038\discretionary{/}{%
}{/}s41467\discretionary{%
}{-}{-}018\discretionary{%
}{-}{-}04608\discretionary{%
}{-}{-}8}}}


\bibitem{adibi2015mobile}
S.~Adibi.
\newblock {\em Mobile Health: A Technology Road Map}, vol.~5.
\newblock Springer, 2015. \href{https://doi.org/10.1007/978-3-319-12817-7}
{doi: {{%
10\hspace{.1pt}\discretionary{.}{%
}{.}\hspace{.4pt}1007\discretionary{/}{%
}{/}978\discretionary{%
}{-}{-}3\discretionary{%
}{-}{-}319\discretionary{%
}{-}{-}12817\discretionary{%
}{-}{-}7}}}


\bibitem{andrienko2013visual}
N.~Andrienko and G.~Andrienko.
\newblock A visual analytics framework for spatio-temporal analysis and modelling.
\newblock {\em Data Min Knowl Discov}, 27:55--83, 2013. \href{https://doi.org/10.1007/s10618-012-0285-7}
{doi: {{%
10\hspace{.1pt}\discretionary{.}{%
}{.}\hspace{.4pt}1007\discretionary{/}{%
}{/}s10618\discretionary{%
}{-}{-}012\discretionary{%
}{-}{-}0285\discretionary{%
}{-}{-}7}}}


\bibitem{andrienko2003exploratory}
N.~Andrienko, G.~Andrienko, and P.~Gatalsky.
\newblock Exploratory spatio-temporal visualization: An analytical review.
\newblock {\em J Visual Lang Comput}, 14(6):503--541, 2003. \href{https://doi.org/10.1016/S1045-926X(03)00046-6}
{doi: {{%
10\hspace{.1pt}\discretionary{.}{%
}{.}\hspace{.4pt}1016\discretionary{/}{%
}{/}S1045\discretionary{%
}{-}{-}926X\discretionary{%
}{(}{(}03\discretionary{)}{%
}{)}00046\discretionary{%
}{-}{-}6}}}


\bibitem{banos2014villalonga}
O.~Banos, R.~Garcia, J.~A. Holgado-Terriza, M.~Damas, H.~Pomares, et~al.
\newblock {mHealthDroid: A} novel framework for agile development of mobile health applications.
\newblock In L.~Pecchia, L.~L. Chen, C.~Nugent, and J.~Bravo, eds., {\em Ambient Assisted Living and Daily Activities}, pp. 91--98. Springer, 2014. \href{https://doi.org/10.1007/978-3-319-13105-4_14}
{doi: {{%
10\hspace{.1pt}\discretionary{.}{%
}{.}\hspace{.4pt}1007\discretionary{/}{%
}{/}978\discretionary{%
}{-}{-}3\discretionary{%
}{-}{-}319\discretionary{%
}{-}{-}13105\discretionary{%
}{-}{-}4\_14}}}


\bibitem{banos2015mhealth}
O.~Banos, C.~Villalonga, R.~García, A.~Saez, M.~Damas, et~al.
\newblock Design, implementation and validation of a novel open framework for agile development of mobile health applications.
\newblock {\em Biomed Eng Online}, 14:S6, 08 2015. \href{https://doi.org/10.1186/1475-925X-14-S2-S6}
{doi: {{%
10\hspace{.1pt}\discretionary{.}{%
}{.}\hspace{.4pt}1186\discretionary{/}{%
}{/}1475\discretionary{%
}{-}{-}925X\discretionary{%
}{-}{-}14\discretionary{%
}{-}{-}S2\discretionary{%
}{-}{-}S6}}}


\bibitem{boileau2020exploring}
P.~Boileau, N.~S. Hejazi, and S.~Dudoit.
\newblock Exploring high-dimensional biological data with sparse contrastive principal component analysis.
\newblock {\em Bioinformatics}, 36(11):3422--3430, 2020. \href{https://doi.org/10.1093/bioinformatics/btaa176}
{doi: {{%
10\hspace{.1pt}\discretionary{.}{%
}{.}\hspace{.4pt}1093\discretionary{/}{%
}{/}bioinformatics\discretionary{/}{%
}{/}btaa176}}}


\bibitem{brandt2023oda}
J.~Brandt, F.~Ciorba, A.~Gentile, M.~Ott, and T.~Wilde.
\newblock Driving {HPC} operations with holistic monitoring and operational data analytics {(Dagstuhl Seminar 23171)}.
\newblock {\em Dagstuhl Reports}, 13(4):98--120, 2023. \href{https://doi.org/10.4230/DagRep.13.4.98}
{doi: {{%
10\hspace{.1pt}\discretionary{.}{%
}{.}\hspace{.4pt}4230\discretionary{/}{%
}{/}DagRep\hspace{.1pt}\discretionary{.}{%
}{.}\hspace{.4pt}13\hspace{.1pt}\discretionary{.}{%
}{.}\hspace{.4pt}4\hspace{.1pt}\discretionary{.}{%
}{.}\hspace{.4pt}98}}}


\bibitem{cao2018voila}
N.~Cao, C.~Lin, Q.~Zhu, Y.-R. Lin, X.~Teng, and X.~Wen.
\newblock Voila: Visual anomaly detection and monitoring with streaming spatiotemporal data.
\newblock {\em IEEE Trans Vis Comput Graph}, 24(1):23--33, 2018. \href{https://doi.org/10.1109/TVCG.2017.2744419}
{doi: {{%
10\hspace{.1pt}\discretionary{.}{%
}{.}\hspace{.4pt}1109\discretionary{/}{%
}{/}TVCG\hspace{.1pt}\discretionary{.}{%
}{.}\hspace{.4pt}2017\hspace{.1pt}\discretionary{.}{%
}{.}\hspace{.4pt}2744419}}}


\bibitem{carroll1970analysis}
J.~D. Carroll and J.-J. Chang.
\newblock Analysis of individual differences in multidimensional scaling via an {N}-way generalization of {``Eckart--Young''} decomposition.
\newblock {\em Psychometrika}, 35(3):283--319, 1970. \href{https://doi.org/10.1007/BF02310791}
{doi: {{%
10\hspace{.1pt}\discretionary{.}{%
}{.}\hspace{.4pt}1007\discretionary{/}{%
}{/}BF02310791}}}


\bibitem{chari2023specious}
T.~Chari and L.~Pachter.
\newblock The specious art of single-cell genomics.
\newblock {\em PLOS Comput Biol}, 19(8):e1011288, 2023.
\newblock 20 pages. \href{https://doi.org/10.1371/journal.pcbi.1011288}
{doi: {{%
10\hspace{.1pt}\discretionary{.}{%
}{.}\hspace{.4pt}1371\discretionary{/}{%
}{/}journal\hspace{.1pt}\discretionary{.}{%
}{.}\hspace{.4pt}pcbi\hspace{.1pt}\discretionary{.}{%
}{.}\hspace{.4pt}1011288}}}


\bibitem{clemmensen2011sparse}
L.~Clemmensen, T.~Hastie, D.~Witten, and B.~Ersb{\o}ll.
\newblock Sparse discriminant analysis.
\newblock {\em Technometrics}, 53(4):406--413, 2011. \href{https://doi.org/10.1198/TECH.2011.08118}
{doi: {{%
10\hspace{.1pt}\discretionary{.}{%
}{.}\hspace{.4pt}1198\discretionary{/}{%
}{/}TECH\hspace{.1pt}\discretionary{.}{%
}{.}\hspace{.4pt}2011\hspace{.1pt}\discretionary{.}{%
}{.}\hspace{.4pt}08118}}}


\bibitem{cunningham2015linear}
J.~P. Cunningham and Z.~Ghahramani.
\newblock Linear dimensionality reduction: Survey, insights, and generalizations.
\newblock {\em J Mach Learn Res}, 16(1):2859--2900, 2015.
\newblock \url{https://jmlr.org/papers/volume16/cunningham15a/cunningham15a.pdf}.

\bibitem{fournet2014contact}
J.~Fournet and A.~Barrat.
\newblock Contact patterns among high school students.
\newblock {\em PLOS One}, 9(9), 2014. \href{https://doi.org/10.1371/journal.pone.0107878}
{doi: {{%
10\hspace{.1pt}\discretionary{.}{%
}{.}\hspace{.4pt}1371\discretionary{/}{%
}{/}journal\hspace{.1pt}\discretionary{.}{%
}{.}\hspace{.4pt}pone\hspace{.1pt}\discretionary{.}{%
}{.}\hspace{.4pt}0107878}}}


\bibitem{fujita2022visual}
K.~Fujita, N.~Sakamoto, T.~Fujiwara, T.~Tsukamoto, and J.~Nonaka.
\newblock A visual analytics method for time-series log data using multiple dimensionality reduction.
\newblock {\em J Adv Simul Sci Eng}, 9(2):206--219, 2022. \href{https://doi.org/10.15748/jasse.9.206}
{doi: {{%
10\hspace{.1pt}\discretionary{.}{%
}{.}\hspace{.4pt}15748\discretionary{/}{%
}{/}jasse\hspace{.1pt}\discretionary{.}{%
}{.}\hspace{.4pt}9\hspace{.1pt}\discretionary{.}{%
}{.}\hspace{.4pt}206}}}


\bibitem{fujiwara2019supporting}
T.~Fujiwara, O.-H. Kwon, and K.-L. Ma.
\newblock Supporting analysis of dimensionality reduction results with contrastive learning.
\newblock {\em IEEE Trans Vis Comput Graph}, 26(1):45--55, 2020. \href{https://doi.org/10.1109/TVCG.2019.2934251}
{doi: {{%
10\hspace{.1pt}\discretionary{.}{%
}{.}\hspace{.4pt}1109\discretionary{/}{%
}{/}TVCG\hspace{.1pt}\discretionary{.}{%
}{.}\hspace{.4pt}2019\hspace{.1pt}\discretionary{.}{%
}{.}\hspace{.4pt}2934251}}}


\bibitem{fujiwara2023contrastive}
T.~Fujiwara and T.-P. Liu.
\newblock Contrastive multiple correspondence analysis ({cMCA}): Using contrastive learning to identify latent subgroups in political parties.
\newblock {\em PLOS ONE}, 18(7):e0287180, 2023. \href{https://doi.org/10.1371/journal.pone.0287180}
{doi: {{%
10\hspace{.1pt}\discretionary{.}{%
}{.}\hspace{.4pt}1371\discretionary{/}{%
}{/}journal\hspace{.1pt}\discretionary{.}{%
}{.}\hspace{.4pt}pone\hspace{.1pt}\discretionary{.}{%
}{.}\hspace{.4pt}0287180}}}


\bibitem{Fujiwara:2021:MultiDR}
T.~Fujiwara, Shilpika, N.~Sakamoto, J.~Nonaka, K.~Yamamoto, and K.-L. Ma.
\newblock A visual analytics framework for reviewing multivariate time-series data with dimensionality reduction.
\newblock {\em IEEE Trans Vis Comput Graph}, 27(2):1601--1611, 2021. \href{https://doi.org/10.1109/TVCG.2020.3028889}
{doi: {{%
10\hspace{.1pt}\discretionary{.}{%
}{.}\hspace{.4pt}1109\discretionary{/}{%
}{/}TVCG\hspace{.1pt}\discretionary{.}{%
}{.}\hspace{.4pt}2020\hspace{.1pt}\discretionary{.}{%
}{.}\hspace{.4pt}3028889}}}


\bibitem{Fujiwara:2022:ULCA}
T.~Fujiwara, X.~Wei, J.~Zhao, and K.-L. Ma.
\newblock Interactive dimensionality reduction for comparative analysis.
\newblock {\em IEEE Trans Vis Comput Graph}, 28(1):758--768, 2022. \href{https://doi.org/10.1109/TVCG.2021.3114807}
{doi: {{%
10\hspace{.1pt}\discretionary{.}{%
}{.}\hspace{.4pt}1109\discretionary{/}{%
}{/}TVCG\hspace{.1pt}\discretionary{.}{%
}{.}\hspace{.4pt}2021\hspace{.1pt}\discretionary{.}{%
}{.}\hspace{.4pt}3114807}}}


\bibitem{ge2016rich}
R.~Ge and J.~Zou.
\newblock Rich component analysis.
\newblock In {\em Proc. ICML}, pp. 1502--1510, 2016.
\newblock \url{https://proceedings.mlr.press/v48/gea16.html}.

\bibitem{gleicher2013explainers}
M.~Gleicher.
\newblock Explainers: Expert explorations with crafted projections.
\newblock {\em IEEE Trans Vis Comput Graph}, 19(12):2042--2051, 2013. \href{https://doi.org/10.1109/TVCG.2013.157}
{doi: {{%
10\hspace{.1pt}\discretionary{.}{%
}{.}\hspace{.4pt}1109\discretionary{/}{%
}{/}TVCG\hspace{.1pt}\discretionary{.}{%
}{.}\hspace{.4pt}2013\hspace{.1pt}\discretionary{.}{%
}{.}\hspace{.4pt}157}}}


\bibitem{golkar2023online}
S.~Golkar, D.~Lipshutz, T.~Tesileanu, and D.~B. Chklovskii.
\newblock An online algorithm for contrastive principal component analysis.
\newblock In {\em Proc. ICASSP}, pp. 1--5. IEEE, 2023. \href{https://doi.org/10.1109/ICASSP49357.2023.10096380}
{doi: {{%
10\hspace{.1pt}\discretionary{.}{%
}{.}\hspace{.4pt}1109\discretionary{/}{%
}{/}ICASSP49357\hspace{.1pt}\discretionary{.}{%
}{.}\hspace{.4pt}2023\hspace{.1pt}\discretionary{.}{%
}{.}\hspace{.4pt}10096380}}}


\bibitem{guo2007regularized}
Y.~Guo, T.~Hastie, and R.~Tibshirani.
\newblock Regularized linear discriminant analysis and its application in microarrays.
\newblock {\em Biostatistics}, 8(1):86--100, 2007. \href{https://doi.org/10.1093/biostatistics/kxj035}
{doi: {{%
10\hspace{.1pt}\discretionary{.}{%
}{.}\hspace{.4pt}1093\discretionary{/}{%
}{/}biostatistics\discretionary{/}{%
}{/}kxj035}}}


\bibitem{hare2015using}
C.~Hare, D.~A. Armstrong, R.~Bakker, R.~Carroll, and K.~T. Poole.
\newblock Using bayesian {Aldrich-McKelvey} scaling to study citizens' ideological preferences and perceptions.
\newblock {\em Am J Polit Sci}, 59(3):759--774, 2015. \href{http://dx.doi.org/10.2139/ssrn.3375435}
{doi: {{%
10\hspace{.1pt}\discretionary{.}{%
}{.}\hspace{.4pt}2139\discretionary{/}{%
}{/}ssrn\hspace{.1pt}\discretionary{.}{%
}{.}\hspace{.4pt}3375435}}}


\bibitem{harshman1970foundations}
R.~A. Harshman.
\newblock Foundations of the {PARAFAC} procedure: Models and conditions for an ``explanatory'' multi-mode factor analysis.
\newblock {\em UCLA Working Papers in Phonetics}, 16:1--84, 1970.

\bibitem{izenman2008modern}
A.~J. Izenman.
\newblock Linear discriminant analysis.
\newblock In {\em Modern Multivariate Statistical Techniques: Regression, Classification, and Manifold Learning}, pp. 237--280. Springer, 2013. \href{https://doi.org/10.1007/978-0-387-78189-1}
{doi: {{%
10\hspace{.1pt}\discretionary{.}{%
}{.}\hspace{.4pt}1007\discretionary{/}{%
}{/}978\discretionary{%
}{-}{-}0\discretionary{%
}{-}{-}387\discretionary{%
}{-}{-}78189\discretionary{%
}{-}{-}1}}}


\bibitem{hyeon2025unveiling}
H.~Jeon, H.~Lee, Y.-H. Kuo, T.~Yang, D.~Archambault, et~al.
\newblock Unveiling high-dimensional backstage: A survey for reliable visual analytics with dimensionality reduction.
\newblock In {\em Proc CHI}. ACM, New York, 2025. \href{https://doi.org/10.1145/3706598.3713551}
{doi: {{%
10\hspace{.1pt}\discretionary{.}{%
}{.}\hspace{.4pt}1145\discretionary{/}{%
}{/}3706598\hspace{.1pt}\discretionary{.}{%
}{.}\hspace{.4pt}3713551}}}


\bibitem{Jolliffe2002Principal}
I.~T. Jolliffe.
\newblock {\em Principal Component Analysis}.
\newblock Springer Series in Statistics. Springer, 2002. \href{https://doi.org/10.1007/b98835}
{doi: {{%
10\hspace{.1pt}\discretionary{.}{%
}{.}\hspace{.4pt}1007\discretionary{/}{%
}{/}b98835}}}


\bibitem{kaya2016high}
O.~Kaya and B.~U{\c{c}}ar.
\newblock High performance parallel algorithms for the {Tucker} decomposition of sparse tensors.
\newblock In {\em Proc. ICPP}, pp. 103--112. IEEE, 2016.

\bibitem{kiers1991hierarchical}
H.~A. Kiers.
\newblock Hierarchical relations among three-way methods.
\newblock {\em Psychometrika}, 56(3):449--470, 1991. \href{https://doi.org/10.1007/BF02294485}
{doi: {{%
10\hspace{.1pt}\discretionary{.}{%
}{.}\hspace{.4pt}1007\discretionary{/}{%
}{/}BF02294485}}}


\bibitem{kiers2000towards}
H.~A. Kiers.
\newblock Towards a standardized notation and terminology in multiway analysis.
\newblock {\em J Chemom}, 14(3):105--122, 2000. \href{https://doi.org/10.1002/1099-128X(200005/06)14:3<105::AID-CEM582>3.0.CO;2-I}
{doi: {{%
10\hspace{.1pt}\discretionary{.}{%
}{.}\hspace{.4pt}1002\discretionary{/}{%
}{/}1099\discretionary{%
}{-}{-}128X\discretionary{%
}{(}{(}200005\discretionary{/}{%
}{/}06\discretionary{)}{%
}{)}14\discretionary{:}{%
}{:}3{\textless}105\discretionary{:}{%
}{:}\discretionary{:}{%
}{:}AID\discretionary{%
}{-}{-}CEM582{\textgreater}3\hspace{.1pt}\discretionary{.}{%
}{.}\hspace{.4pt}0\hspace{.1pt}\discretionary{.}{%
}{.}\hspace{.4pt}CO\discretionary{;}{%
}{;}2\discretionary{%
}{-}{-}I}}}


\bibitem{kolda2009tensor}
T.~G. Kolda and B.~W. Bader.
\newblock Tensor decompositions and applications.
\newblock {\em SIAM Rev}, 51(3):455--500, 2009. \href{https://doi.org/10.1137/07070111X}
{doi: {{%
10\hspace{.1pt}\discretionary{.}{%
}{.}\hspace{.4pt}1137\discretionary{/}{%
}{/}07070111X}}}


\bibitem{JMLR:v20:18-277}
J.~Kossaifi, Y.~Panagakis, A.~Anandkumar, and M.~Pantic.
\newblock {TensorLy}: Tensor learning in python.
\newblock {\em J Mach Learn Res}, 20(26):1--6, 2019.
\newblock \url{http://jmlr.org/papers/v20/18-277.html}.

\bibitem{lai2013sparse}
Z.~Lai, Y.~Xu, J.~Yang, J.~Tang, and D.~Zhang.
\newblock Sparse tensor discriminant analysis.
\newblock {\em IEEE Trans Image Process}, 22(10):3904--3915, 2013. \href{https://doi.org/10.1109/TIP.2013.2264678}
{doi: {{%
10\hspace{.1pt}\discretionary{.}{%
}{.}\hspace{.4pt}1109\discretionary{/}{%
}{/}TIP\hspace{.1pt}\discretionary{.}{%
}{.}\hspace{.4pt}2013\hspace{.1pt}\discretionary{.}{%
}{.}\hspace{.4pt}2264678}}}


\bibitem{crl}
P.~H. Le-Khac, G.~Healy, and A.~F. Smeaton.
\newblock Contrastive representation learning: A framework and review.
\newblock {\em IEEE Access}, 8:193907--193934, 2020. \href{https://doi.org/10.1109/ACCESS.2020.3031549}
{doi: {{%
10\hspace{.1pt}\discretionary{.}{%
}{.}\hspace{.4pt}1109\discretionary{/}{%
}{/}ACCESS\hspace{.1pt}\discretionary{.}{%
}{.}\hspace{.4pt}2020\hspace{.1pt}\discretionary{.}{%
}{.}\hspace{.4pt}3031549}}}


\bibitem{Liu:2019:TPFlow}
D.~Liu, P.~Xu, and L.~Ren.
\newblock {TPFlow}: Progressive partition and multidimensional pattern extraction for large-scale spatio-temporal data analysis.
\newblock {\em IEEE Trans Vis Comput Graph}, 25(1):1--11, 2019. \href{https://doi.org/10.1109/TVCG.2018.2865018}
{doi: {{%
10\hspace{.1pt}\discretionary{.}{%
}{.}\hspace{.4pt}1109\discretionary{/}{%
}{/}TVCG\hspace{.1pt}\discretionary{.}{%
}{.}\hspace{.4pt}2018\hspace{.1pt}\discretionary{.}{%
}{.}\hspace{.4pt}2865018}}}


\bibitem{lu2011survey}
H.~Lu, K.~N. Plataniotis, and A.~N. Venetsanopoulos.
\newblock A survey of multilinear subspace learning for tensor data.
\newblock {\em Pattern Recognit}, 44(7):1540--1551, 2011. \href{https://doi.org/10.1016/j.patcog.2011.01.004}
{doi: {{%
10\hspace{.1pt}\discretionary{.}{%
}{.}\hspace{.4pt}1016\discretionary{/}{%
}{/}j\hspace{.1pt}\discretionary{.}{%
}{.}\hspace{.4pt}patcog\hspace{.1pt}\discretionary{.}{%
}{.}\hspace{.4pt}2011\hspace{.1pt}\discretionary{.}{%
}{.}\hspace{.4pt}01\hspace{.1pt}\discretionary{.}{%
}{.}\hspace{.4pt}004}}}


\bibitem{lu2024geometric}
R.-S. Lu, S.-H. Wang, and S.-Y. Huang.
\newblock A geometric algorithm for contrastive principal component analysis in high dimension.
\newblock {\em J Comput Graph Stat}, pp. 1--8, 2024. \href{https://doi.org/10.1080/10618600.2023.2289542}
{doi: {{%
10\hspace{.1pt}\discretionary{.}{%
}{.}\hspace{.4pt}1080\discretionary{/}{%
}{/}10618600\hspace{.1pt}\discretionary{.}{%
}{.}\hspace{.4pt}2023\hspace{.1pt}\discretionary{.}{%
}{.}\hspace{.4pt}2289542}}}


\bibitem{mcinnes2018umap}
L.~McInnes, J.~Healy, and J.~Melville.
\newblock {UMAP}: Uniform manifold approximation and projection for dimension reduction.
\newblock {\em arXiv:1802.03426}, 2018. \href{https://doi.org/10.48550/arXiv.1802.03426}
{doi: {{%
10\hspace{.1pt}\discretionary{.}{%
}{.}\hspace{.4pt}48550\discretionary{/}{%
}{/}arXiv\hspace{.1pt}\discretionary{.}{%
}{.}\hspace{.4pt}1802\hspace{.1pt}\discretionary{.}{%
}{.}\hspace{.4pt}03426}}}


\bibitem{miyazaki2012overview}
H.~Miyazaki, Y.~Kusano, N.~Shinjou, F.~Shoji, M.~Yokokawa, and T.~Watanabe.
\newblock Overview of the {K} computer system.
\newblock {\em Fujitsu Sci Tech J}, 48(3):255--265, 2012.

\bibitem{oseledets2011tensor}
I.~V. Oseledets.
\newblock Tensor-train decomposition.
\newblock {\em SIAM J Sci Comput}, 33(5):2295--2317, 2011. \href{https://doi.org/10.1137/090752286}
{doi: {{%
10\hspace{.1pt}\discretionary{.}{%
}{.}\hspace{.4pt}1137\discretionary{/}{%
}{/}090752286}}}


\bibitem{ott2020oda}
M.~Ott, W.~Shin, N.~Bourassa, T.~Wilde, S.~Ceballos, et~al.
\newblock Global experiences with {HPC} operational data measurement, collection and analysis.
\newblock In {\em Proc. CLUSTER}, pp. 499--508, 2020. \href{https://doi.org/10.1109/CLUSTER49012.2020.00071}
{doi: {{%
10\hspace{.1pt}\discretionary{.}{%
}{.}\hspace{.4pt}1109\discretionary{/}{%
}{/}CLUSTER49012\hspace{.1pt}\discretionary{.}{%
}{.}\hspace{.4pt}2020\hspace{.1pt}\discretionary{.}{%
}{.}\hspace{.4pt}00071}}}


\bibitem{pajarola2021tensor}
R.~Pajarola, S.~K. Suter, R.~Ballester-Ripoll, and H.~Yang.
\newblock Tensor approximation for multidimensional and multivariate data.
\newblock In {\em Anisotropy Across Fields and Scales}, pp. 73--98. Springer, 2021. \href{https://doi.org/10.1007/978-3-030-56215-1_4}
{doi: {{%
10\hspace{.1pt}\discretionary{.}{%
}{.}\hspace{.4pt}1007\discretionary{/}{%
}{/}978\discretionary{%
}{-}{-}3\discretionary{%
}{-}{-}030\discretionary{%
}{-}{-}56215\discretionary{%
}{-}{-}1\_4}}}


\bibitem{dash}
{Plotly}.
\newblock {Dash} {Python} user guide.
\newblock \url{https://dash.plotly.com/}, 2024.
\newblock Accessed: 2024-09-22.

\bibitem{rabanser2017introduction}
S.~Rabanser, O.~Shchur, and S.~G{\"u}nnemann.
\newblock Introduction to tensor decompositions and their applications in machine learning.
\newblock {\em arXiv:1711.10781}, 2017. \href{https://doi.org/10.48550/arXiv.1711.1078}
{doi: {{%
10\hspace{.1pt}\discretionary{.}{%
}{.}\hspace{.4pt}48550\discretionary{/}{%
}{/}arXiv\hspace{.1pt}\discretionary{.}{%
}{.}\hspace{.4pt}1711\hspace{.1pt}\discretionary{.}{%
}{.}\hspace{.4pt}1078}}}


\bibitem{salloum2022cpca++}
R.~Salloum and C.-C.~J. Kuo.
\newblock {cPCA++: An} efficient method for contrastive feature learning.
\newblock {\em Pattern Recognit}, 124:108378, 2022. \href{https://doi.org/10.1016/j.patcog.2021.108378}
{doi: {{%
10\hspace{.1pt}\discretionary{.}{%
}{.}\hspace{.4pt}1016\discretionary{/}{%
}{/}j\hspace{.1pt}\discretionary{.}{%
}{.}\hspace{.4pt}patcog\hspace{.1pt}\discretionary{.}{%
}{.}\hspace{.4pt}2021\hspace{.1pt}\discretionary{.}{%
}{.}\hspace{.4pt}108378}}}


\bibitem{tao2007general}
D.~Tao, X.~Li, X.~Wu, and S.~J. Maybank.
\newblock General tensor discriminant analysis and {Gabor} features for gait recognition.
\newblock {\em IEEE Trans Pattern Anal Mach Intell}, 29(10):1700--1715, 2007. \href{https://doi.org/10.1109/TPAMI.2007.1096}
{doi: {{%
10\hspace{.1pt}\discretionary{.}{%
}{.}\hspace{.4pt}1109\discretionary{/}{%
}{/}TPAMI\hspace{.1pt}\discretionary{.}{%
}{.}\hspace{.4pt}2007\hspace{.1pt}\discretionary{.}{%
}{.}\hspace{.4pt}1096}}}


\bibitem{tenenbaum2000global}
J.~B. Tenenbaum, V.~d. Silva, and J.~C. Langford.
\newblock A global geometric framework for nonlinear dimensionality reduction.
\newblock {\em Science}, 290(5500):2319--2323, 2000. \href{https://doi.org/10.1126/science.290.5500.2319}
{doi: {{%
10\hspace{.1pt}\discretionary{.}{%
}{.}\hspace{.4pt}1126\discretionary{/}{%
}{/}science\hspace{.1pt}\discretionary{.}{%
}{.}\hspace{.4pt}290\hspace{.1pt}\discretionary{.}{%
}{.}\hspace{.4pt}5500\hspace{.1pt}\discretionary{.}{%
}{.}\hspace{.4pt}2319}}}


\bibitem{timmerman2000three}
M.~E. Timmerman and H.~A. Kiers.
\newblock Three-mode principal components analysis: Choosing the numbers of components and sensitivity to local optima.
\newblock {\em Br J Stat Psychol}, 53(1):1--16, 2000. \href{https://doi.org/10.1348/000711000159132}
{doi: {{%
10\hspace{.1pt}\discretionary{.}{%
}{.}\hspace{.4pt}1348\discretionary{/}{%
}{/}000711000159132}}}


\bibitem{townsend2016pymanopt}
J.~Townsend, N.~Koep, and S.~Weichwald.
\newblock Pymanopt: A {Python} toolbox for optimization on manifolds using automatic differentiation.
\newblock {\em J Mach Learn Res}, 17(137):1--5, 2016.
\newblock \url{https://jmlr.org/papers/v17/16-177.html}.

\bibitem{tucker1966some}
L.~R. Tucker.
\newblock Some mathematical notes on three-mode factor analysis.
\newblock {\em Psychometrika}, 31(3):279--311, 1966. \href{https://doi.org/10.1007/BF02289464}
{doi: {{%
10\hspace{.1pt}\discretionary{.}{%
}{.}\hspace{.4pt}1007\discretionary{/}{%
}{/}BF02289464}}}


\bibitem{airdata}
{US Environmental Protection Agency}.
\newblock {Air Data}: Air quality data collected at outdoor monitors across the {US}.
\newblock \url{https://www.epa.gov/outdoor-air-quality-data}, 2019.
\newblock Accessed: 2024-09-23.

\bibitem{van2008visualizing}
L.~van~der Maaten and G.~Hinton.
\newblock Visualizing data using {t-SNE}.
\newblock {\em J Mach Learn Res}, 9(11), 2008.
\newblock \url{https://jmlr.org/papers/v9/vandermaaten08a.html}.

\bibitem{van2009dimensionality}
L.~van~der Maaten, E.~Postma, and J.~van~den Herik.
\newblock Dimensionality reduction: A comparative review.
\newblock Technical Report TiCC-TR 2009-005, Tilburg University Technical Report, 2009.
\newblock 36 pages, \url{https://lvdmaaten.github.io/publications/papers/TR_Dimensionality_Reduction_Review_2009.pdf}.

\bibitem{virtanen2020scipy}
P.~Virtanen, R.~Gommers, T.~E. Oliphant, M.~Haberland, T.~Reddy, et~al.
\newblock {SciPy} 1.0: Fundamental algorithms for scientific computing in {Python}.
\newblock {\em Nat Methods}, 17:261--272, 2020. \href{https://doi.org/10.1038/s41592-019-0686-2}
{doi: {{%
10\hspace{.1pt}\discretionary{.}{%
}{.}\hspace{.4pt}1038\discretionary{/}{%
}{/}s41592\discretionary{%
}{-}{-}019\discretionary{%
}{-}{-}0686\discretionary{%
}{-}{-}2}}}


\bibitem{wang2021understanding}
Y.~Wang, H.~Huang, C.~Rudin, and Y.~Shaposhnik.
\newblock Understanding how dimension reduction tools work: An empirical approach to deciphering {t-SNE, UMAP, TriMAP, and PaCMAP} for data visualization.
\newblock {\em J Mach Learn Res}, 22(201):1--73, 2021.
\newblock \url{https://www.jmlr.org/papers/v22/20-1061.html}.

\bibitem{wang2022iot}
Z.~Wang, H.~Xiong, J.~Zhang, S.~Yang, M.~Boukhechba, et~al.
\newblock From personalized medicine to population health: A survey of mhealth sensing techniques.
\newblock {\em IEEE Internet Things J}, 9(17):15413--15434, 2022. \href{https://doi.org/10.1109/JIOT.2022.3161046}
{doi: {{%
10\hspace{.1pt}\discretionary{.}{%
}{.}\hspace{.4pt}1109\discretionary{/}{%
}{/}JIOT\hspace{.1pt}\discretionary{.}{%
}{.}\hspace{.4pt}2022\hspace{.1pt}\discretionary{.}{%
}{.}\hspace{.4pt}3161046}}}


\bibitem{wen2018robust}
J.~Wen, X.~Fang, J.~Cui, L.~Fei, K.~Yan, et~al.
\newblock Robust sparse linear discriminant analysis.
\newblock {\em IEEE Trans Circ Syst Video Tech}, 29(2):390--403, 2018. \href{https://doi.org/10.1109/TCSVT.2018.2799214}
{doi: {{%
10\hspace{.1pt}\discretionary{.}{%
}{.}\hspace{.4pt}1109\discretionary{/}{%
}{/}TCSVT\hspace{.1pt}\discretionary{.}{%
}{.}\hspace{.4pt}2018\hspace{.1pt}\discretionary{.}{%
}{.}\hspace{.4pt}2799214}}}


\bibitem{yan2005discriminant}
S.~Yan, D.~Xu, Q.~Yang, L.~Zhang, X.~Tang, and H.-J. Zhang.
\newblock Discriminant analysis with tensor representation.
\newblock In {\em Proc. CVPR}, vol.~1, pp. 526--532. IEEE, 2005. \href{https://doi.org/10.1109/CVPR.2005.131}
{doi: {{%
10\hspace{.1pt}\discretionary{.}{%
}{.}\hspace{.4pt}1109\discretionary{/}{%
}{/}CVPR\hspace{.1pt}\discretionary{.}{%
}{.}\hspace{.4pt}2005\hspace{.1pt}\discretionary{.}{%
}{.}\hspace{.4pt}131}}}


\bibitem{yang2004two}
J.~Yang, D.~Zhang, A.~F. Frangi, and J.-y. Yang.
\newblock {Two-dimensional PCA}: A new approach to appearance-based face representation and recognition.
\newblock {\em IEEE Trans Pattern Anal Mach Intell}, 26(1):131--137, 2004. \href{https://doi.org/10.1109/TPAMI.2004.1261097}
{doi: {{%
10\hspace{.1pt}\discretionary{.}{%
}{.}\hspace{.4pt}1109\discretionary{/}{%
}{/}TPAMI\hspace{.1pt}\discretionary{.}{%
}{.}\hspace{.4pt}2004\hspace{.1pt}\discretionary{.}{%
}{.}\hspace{.4pt}1261097}}}


\bibitem{yang2005two}
J.~Yang, D.~Zhang, X.~Yong, and J.-y. Yang.
\newblock Two-dimensional discriminant transform for face recognition.
\newblock {\em Pattern Recognit}, 38(7):1125--1129, 2005. \href{https://doi.org/10.1016/j.patcog.2004.11.019}
{doi: {{%
10\hspace{.1pt}\discretionary{.}{%
}{.}\hspace{.4pt}1016\discretionary{/}{%
}{/}j\hspace{.1pt}\discretionary{.}{%
}{.}\hspace{.4pt}patcog\hspace{.1pt}\discretionary{.}{%
}{.}\hspace{.4pt}2004\hspace{.1pt}\discretionary{.}{%
}{.}\hspace{.4pt}11\hspace{.1pt}\discretionary{.}{%
}{.}\hspace{.4pt}019}}}


\bibitem{yasir2015comparison}
M.~Yasir, E.~Angelakis, F.~Bibi, E.~Azhar, D.~Bachar, et~al.
\newblock Comparison of the gut microbiota of people in {France} and {Saudi Arabia}.
\newblock {\em Nutr Diabetes}, 5(4):e153--e153, 2015. \href{https://doi.org/10.1038/nutd.2015.3}
{doi: {{%
10\hspace{.1pt}\discretionary{.}{%
}{.}\hspace{.4pt}1038\discretionary{/}{%
}{/}nutd\hspace{.1pt}\discretionary{.}{%
}{.}\hspace{.4pt}2015\hspace{.1pt}\discretionary{.}{%
}{.}\hspace{.4pt}3}}}


\bibitem{zou2013contrastive}
J.~Y. Zou, D.~J. Hsu, D.~C. Parkes, and R.~P. Adams.
\newblock Contrastive learning using spectral methods.
\newblock In {\em Proc. NIPS}, pp. 2238--2246, 2013.
\newblock \url{https://papers.nips.cc/paper_files/paper/2013/hash/36a16a2505369e0c922b6ea7a23a56d2-Abstract.html}.

\end{thebibliography}








\end{document}